\documentclass[11pt]{article}
\pdfoutput=1  
\usepackage{graphicx}
\usepackage{appendix}
\usepackage{amssymb}
\usepackage{amsmath}
\usepackage{amsfonts}
\usepackage{upgreek}
\usepackage{latexsym}
\usepackage{slashed,upgreek,amscd,cancel,tensor}
\usepackage{adjustbox}
\usepackage{epsfig}
\usepackage{cite}
\usepackage{hyperref}
\usepackage{amsfonts}
\usepackage{booktabs}
\usepackage{siunitx}
\usepackage{caption}
\usepackage{float}
\usepackage{color}
\usepackage{tikz-feynman}
\usepackage{subfig}
\usepackage{physics}
\usepackage{array}
\usepackage{soul}
\usepackage{xcolor}

\numberwithin{equation}{section}

\definecolor{MyBlue}{rgb}{0.15,0.15,0.70}

\hypersetup{
colorlinks=true,
citecolor=MyBlue,
linkcolor=MyBlue,
urlcolor=MyBlue
}

\input epsf
\newcommand{\Mp}{M_{\rm Pl}}
\newcommand{\aaa}{\alpha}
\newcommand{\bbb}{\beta}

\newcommand{\be}{\begin{equation}}
\newcommand{\ee}{\end{equation}}
\newcommand{\beq}{\begin{equation}}
\newcommand{\eeq}{\end{equation}}
\newcommand{\bea}{\begin{eqnarray}}
\newcommand{\eea}{\end{eqnarray}}

\def\dkmu2{\delta K_{\mu \nu}\delta K^{\mu \nu}}
\def\pmu2{  \phi_{\mu \nu}\phi^{\mu \nu}}

\renewcommand\[{\left[}

\newcommand\ees{\end{eqnarray}}
\newcommand\bees{\begin{eqnarray}}

\begin{document}

\begin{center}
\LARGE{\bf Effective field theory for gravitational radiation\\ 
in scalar-tensor gravity}
\\[1cm]

\large{Adrien Kuntz$^{\rm a}$,  Federico Piazza$^{\rm a}$ and Filippo Vernizzi$^{\rm b}$}
\\[0.5cm]

\small{
\textit{$^{\rm a}$
Aix Marseille Univ, Universit\'e de Toulon, CNRS, CPT, Marseille, France}}
\vspace{.2cm}

\small{
\textit{$^{\rm b}$ Institut de physique th\' eorique, Universit\'e  Paris Saclay, CEA, CNRS \\ [0.05cm]
 91191 Gif-sur-Yvette, France}}

\vspace{.2cm}

\end{center}

\begin{abstract} 
A light scalar degree of freedom, as the one possibly responsible for the accelerated expansion of the Universe, could leave observable traces in the inspiral gravitational wave signal of  binary systems. In order to study these effects, we extend the effective field theory formalism of Goldberger and Rothstein to minimal scalar-tensor theories of gravity. This class of models is still very broad, because the couplings of the scalar to matter are far less constrained than those a massless spin-2 field. In most of the paper we focus on conformal couplings. Using the effective field theory approach, we discuss the emergence of violations of the strong equivalence principle even in models  that exhibit universality of free fall at the microscopic level. Our results on  the conservative dynamics of the binary and its power emitted in gravitational and scalar radiation agree with those obtained with the standard post-Newtonian formalism. We also compare them to more recent work. Finally, we  discuss the implications of considering a disformal coupling to matter. 

\end{abstract}

\newpage
\tableofcontents
\vspace{0.5cm} 

\newpage

\section{Introduction}
The direct detection of gravitational waves (GW)~\cite{Abbott:2016blz}, while opening an entire new window on astrophysics and cosmology, is providing a precious direct test of general relativity (GR). One of the main reasons why one could feel legitimated to put GR into question is the latest acceleration of the Universe, a phenomenon whose cause is broadly called {\it dark energy}, which contemplates some modification of gravity among its possible explanations. Of course, we are talking about very different scales here: cosmic acceleration is active at  Hubble scales, i.e.~at distances much larger than the size of any gravitational wave emitter. 

On the other hand, the dynamics of a massless spin-2 field is extremely constrained on theoretical grounds. According to our present understanding of effective field theory (EFT), modifying GR necessarily implies introducing new degrees of freedom besides the massless spin-2 field. This could have several effects on GW observables. First, the
new degree of freedom may exchange a fifth force, changing the conservative and dissipative dynamics of the inspiral binaries emitting GW. Second, the
new propagating degree of freedom opens new channels of radiation emission that can be observed as extra polarizations. Third, if this new field develops a non-trivial cosmological background value, it can induce Lorentz-breaking effects on the GW such as modifying their dispersion relation (see e.g.~\cite{Yunes:2016jcc}). 
The most common of these types of effects is inducing a non-luminal GW speed \cite{Gleyzes:2013ooa,Jimenez:2015bwa,Lombriser:2015sxa,Bettoni:2016mij}. The recent  detection of the neutron star merger GW170817 both in the gravitational and electromagnetic channels~\cite{the_ligo_scientific_collaboration_gw170817:_2017}, 
has constrained such an effect to the impressive level of $10^{-15}$, killing a large set of modified gravity models~\cite{Creminelli:2017sry,Ezquiaga:2017ekz,Baker:2017hug,Sakstein:2017xjx}.  The consequences of this event on the Vainshtein mechanism in scalar-tensor theories  have
been discussed in  \cite{Crisostomi:2017lbg,Langlois:2017dyl,dima_vainshtein_2018}. 
Finally,   GW may be damped \cite{Deffayet:2007kf,Calabrese:2016bnu,Visinelli:2017bny,Amendola:2017ovw,Belgacem:2017ihm} or decay into dark energy fluctuations \cite{Creminelli:2018xsv}. See also \cite{Ezquiaga:2018btd} for a review on the impact of present and future GW observations on modified gravity.

In this work we study gravitational and scalar waves emission of a binary system in a post-Newtonian expansion (see e.g.~\cite{Blanchet:2013haa} and references therein), 
for theories that modify gravity by the addition of a light scalar degree of freedom. To this aim, we extend the formalism of Non-Relativistic General Relativity (NRGR), developed by Goldberger and Rothstein  \cite{goldberger_effective_2006,Goldberger:2007hy} (see also \cite{Cardoso:2008gn,Galley:2009px}, the nice reviews~\cite{Foffa:2013qca,porto_effective_2016,Levi:2018nxp} and the interesting extension discussed in \cite{Endlich:2017tqa}), to include an extra light  scalar besides the massless graviton. 
After having presented the general formalism, we highlight how violations of the strong Equivalence Principle, as first derived by Nordtvedt \cite{Nordtvedt:1968qs}, emerge in this approach. Moreover, we derive the emitted gravitational and scalar power and the observable waveform of the two tensor polarizations and the scalar one.
Although we use different methods to obtain them, our results 
overlap with some classic works of Damour and Esposito-Far\`{e}se, where the standard Post-Newtonian formalism is applied to (multi) scalar-tensor theories~\cite{damour_tensor-multi-scalar_1992, Damour:1995kt}\footnote{It should be noted that \cite{Damour:1995kt} also uses a field theory language (in position space) in the conservative sector of the dynamics. Our approach complements this work by enforcing the power-counting rules directly at the level of the diagrammatic expansion.}. When possible, we compare with the more recent Ref.~\cite{Huang:2018pbu}, where NRGR is used to study the effects of a light axion. 
For recent developments on the two-body problem 
in scalar-tensor theories see also \cite{Julie:2017ucp,Julie:2017pkb}, where the conservative dynamics  have been extended to the Effective-One-Body framework at 2PN order, and \cite{Bernard:2018hta, Bernard:2018ivi} where the equations of motion have been derived  up to 3PN order. See  \cite{Lang:2013fna,Lang:2014osa} for the tensor and scalar waveform calculations, respectively at 2PN and 1.5PN order.

To describe the binary system coupled by gravity and the emitted gravitational wave we use the action
\be
\label{totaction}
S  =  S_{\rm grav} + S_{\rm pp} \;,
\ee
where $S_{\rm grav}$ governs the dynamics of the gravitational and scalar degrees of freedom and is given by
\begin{equation} \label{firstaction}
S_{\rm grav} = \:  \int d^4x \sqrt{-g} \left(\frac{\Mp^2}{2} R - \frac{1}{2}  g^{\mu \nu}\partial_\mu \phi \partial_\nu \phi \right)\, ,
\end{equation}
while $S_{\rm pp}$ is the point-particle action describing the motion of the two inspiralling objects, labelled by $A=1,2$, 
\begin{equation} 
\label{eq:action}
S_{\rm pp} = -\sum _{A=1,2} m_A \int d\tau_A \left[ 1 - \aaa_A \frac{\phi}{\Mp}  - \bbb_A \left(\frac{\phi}{\Mp}\right)^2 +\dots \right] \, ,
\end{equation}
where $d\tau_A^2 = g_{\mu \nu} d  x_A^\mu d x_A^\nu$ is the proper time of each object. The term in square brackets can be seen as originating from a Taylor expansion of a ``$\phi$-dependent mass'' $m_A(\phi)$~\cite{Eardley1975ApJ}. We now discuss why we restrict to this action and some ways to extend it.

\subsection{The gravitational action}

While trying to extend the NRGR formalism to dark energy/scalar field models we immediately encounter an obstruction. 
It is well known that, to pass Solar System tests, modified gravity theories   display {\it screening} mechanisms that make the scalar interactions weaker in high density environments. For instance, most theories belonging to the Galileon/Horndeski~\cite{Nicolis:2008in,Horndeski:1974wa,Deffayet:2009mn} and ``beyond Horndeski"~\cite{Zumalacarregui:2013pma,Gleyzes:2014dya,Gleyzes:2014qga,Langlois:2015cwa,BenAchour:2016fzp} classes  have a rich structure of non-linear terms in their Lagrangians that become more important close to the sources, thereby screening the effects of the scalar fluctuations \cite{Vainshtein:1972sx,Babichev:2013usa}. While it is legitimate to neglect such non-linearities on the largest cosmological scales, they are expected to play a major role in the vicinity of the binaries, causing a breakdown of the perturbative expansion that we use in this paper.

Another, related, complication in dealing with dark energy is represented by the spontaneously breaking of Lorentz symmetry. 
A time-evolving background scalar field allows, in the action that governs cosmological perturbations, all sorts of terms that break boosts and time-translations. This is made particularly explicit in the {\it EFT approach} to inflation \cite{Creminelli:2006xe,Cheung:2007st} and dark energy~\cite{Creminelli:2008wc,Gubitosi:2012hu,Bloomfield:2012ff,Piazza:2013coa}. Let us consider, as an example, the Nambu-Goldstone boson of some broken $U(1)$ symmetry in Minkowski space, 
\begin{equation}
\label{eq:kmoufla}
{\cal L} = - \partial_\mu  \phi \partial^\mu \phi + \frac{(\partial_\mu  \phi \partial^\mu \phi)^2}{\Lambda_*^4} + \dots \, , 
\end{equation}
where $\Lambda_*$ is a mass scale. When this theory is expanded around a homogeneous background configuration $\phi (t)$, which is always a solution in Minkowski, the fluctuations of the field develop a speed of propagation $c_s\neq1$---a very tangible sign of spontaneous breaking of Lorentz symmetry already at quadratic order. 
At the same time, the non-linear terms suppressed by the scale $\Lambda_*$  become important close to the source, giving rise, in this case, to  the ``$k$-mouflage" mechanism~\cite{Babichev:2009ee}, which is a variant of the screening mechanisms discussed above. 

As a full treatment of these  non-linearities is beyond the scope of this paper (see however~\cite{Chu:2012kz,deRham:2012fw,deRham:2012fg,Dar:2018dra} for interesting steps forward in this direction), we focus on a very standard scalar-tensor action, i.e.~eq.~\eqref{firstaction}. 
This action is in the so-called ``Einstein frame'' form: possible non minimal couplings between the scalar and the metric fields have been reabsorbed with a field redefinition of the metric and transferred to the matter sector, i.e., to the point-particle action. Moreover, motivated by the dark energy role of the scalar field, we assume it to be effectively massless. The case of a massive axion has been studied in the effective field theory approach in \cite{Huang:2018pbu}.

\subsection{The point-particle action}

Let us now motivate the point-particle action  \eqref{eq:action}. The size of the objects that are orbiting around each other represents our UV scale, so they can be effectively described as point particles. The couplings of a massless spin-2 field are famously constrained by gauge invariance \cite{Weinberg:1964ew}. As a result, the presence of a scalar provides us with a  richer structure of possible point-particle couplings.

For most of the paper we focus on {\it conformal couplings}, obtained by assuming that matter couples to the gravitational metric multiplied by a general function of the scalar field.
We have included up to terms quadratic in $\phi$ because, as we will see, higher-order terms become important only at higher order in the Post-Newtonian expansion. Moreover, we have allowed different scalar couplings for different particles because such couplings are not protected against renormalization.

For example, let us consider  the actual field theory describing the matter inside the object. Such a ``UV model" might well enjoy a universal scalar coupling of the type 
\begin{equation}
S_{\rm UV} \supset \int d^4x\  \alpha \ \frac{\phi}{M_P} T_{\rm m},
\end{equation}
$T_{\rm m}$ being the trace of the energy momentum tensor of the matter fields. Such a universal coupling is indeed radiatively stable under corrections coming from the sole matter sector (see e.g.~\cite{Hui:2010dn,Nitti:2012ev}). However, the matching into the EFT point-particle action~\eqref{eq:action}  inevitably contains details about the actual shape and density of the body under consideration. For example, as a body becomes more and more self-gravitating, its scalar charge  decreases, down to the point of disappearing when it becomes a black hole (see e.g. the nice discussion in~\cite{Hui:2009kc}). Equivalently, if we started with a universal EFT model~\eqref{eq:action} with some bare mass parameters $m_{\text{bare},A}$ and universal couplings $\alpha_{\text{bare},1} = \alpha_{\text{bare},2} = \alpha_\text{bare}$ (and $\beta_{\text{bare},1} = \beta_{\text{bare},2} = \beta_\text{bare}$),  we can make the corrections to the mass finite by imposing a hard cutoff $\Lambda$ in momentum space, which roughly corresponds to considering a body of size $\Lambda^{-1}$. We get
\begin{equation}
m_A (\Lambda) \ = \  m_{\text{bare},A} + \delta m_A(\Lambda)\, ,
\end{equation}
where $\delta m_A(\Lambda)$ represents the (negative) gravitational energy of the body. The explicit calculation is done in Sec.~\ref{sec2.2}. As we show there, the scalar charge of the body renormalizes in a way that is not universal but actually depends on $\delta m_A(\Lambda)$ (see equation~\eqref{eq:deltaalpha}). 
This result is often stated by saying that a scalar fifth force can satisfy the weak equivalence principle (universality of the free fall for test particles) but not the strong one (universality of the free fall for bodies of non-negligible gravitational self-energy). It is believed that the latter is satisfied only by a purely metric theory as GR~\cite{Will:1993ns}.
It is therefore important to allow different scalar couplings for different objects at the level of the EFT, with the understanding that, in most cases, such a charge is zero for a black hole. 

In App.~\ref{app} we attempt a first systematic discussion of such couplings. In Sec.~\ref{Sec:dis}
we present a second example of scalar-point-particle coupling that can be extracted from the very general action~\eqref{line13}-\eqref{line15}, i.e.~the {\it disformal coupling}. This corresponds to a standard metric coupling to the point-particle trajectory $x^\mu(t)$,
\begin{equation}
S_{\rm pp} = - m \int d t \sqrt{- \tilde g_{\mu \nu} v^\mu v^\nu } \, ,
\end{equation}  
where 
\be
v^\mu \equiv \frac{d x^\mu}{d t} \;
\ee 
is the {\it four}-velocity vector and
the metric $\tilde g_{\mu \nu}$ is defined as 
\begin{equation}
\label{disformal}
\tilde{g}_{\mu \nu} = A(\phi) g_{\mu \nu} + \frac{1}{\Lambda_*^4} \partial_\mu \phi \partial_\nu \phi\, .
\end{equation}
One interesting aspect of such a coupling is that it exhibits {\it non-linearities} at small distances, analogous to those responsible for screening effects in $k$-mouflage theories. 
In the absence of non-linear terms in the field Lagrangian this seems hardly possible. After all, the vacuum field equation that we get from~\eqref{firstaction} is still
$\Box \phi =0$, which means that stationary field configurations must display the usual $1/r$ behavior even arbitrarily close to the source. On the opposite, in theories with  screening, the scalar field profile smoothens  as we get closer to the origin---that is when the non-linear terms in the Lagrangian take over. 

The point is that, in the presence of a disformal coupling, the one-body static solution does not capture some velocity-dependent non-linear features, which appear only at the level of the two-body interactions. As detailed in Sec.~\ref{Sec:dis}, there is a typical distance $r_* \simeq ( \alpha m /\Mp)^{1/2} / \Lambda_*$, at which the diagrammatic expansion breaks down because higher-order diagrams become more important than lower-order ones.  Such a non-linear radius can be parametrically larger than the Schwarzschild one and, for values of  $\Lambda_*$ of cosmological interest, i.e.~$\Lambda_* \simeq ( \Mp H_0)^{1/2}$, can end up being of the same order as the $k$-mouflage radius. For example, for an object of the same mass as the sun, $r_* \sim$ 1 parsec.

\subsection{Outline}

The plan of the paper is as follows. In Sec.~\ref{sec3} we extend the  formalism of Goldberger and Rothstein  to modified gravity and  introduce the Non-Relativistic Scalar-Tensor formalism. This part should be understood as a toolbox, which provides all the necessary material that is needed in order to reproduce the results obtained in the other sections. Section \ref{sec2.2} deals with the renormalization of masses and charges due to the gravitational energy of the bodies that we consider, and Sec.~\ref{sec:EIH_Lagrangian} presents the first relativistic correction to the two-body Lagrangian. The corresponding Lagrangian in GR is the famous EIH Lagrangian, named after Einstein, Infeld and Hoffmann. In Sec.~\ref{sec:couplings} we explain the multipole expansion of radiative fields at the level of the action, and we use it to describe in Sec.~\ref{sec:dissipative_dynamics} the dissipative dynamics of the system. Before concluding, we comment on the effects of a disformal coupling in Sec.~\ref{Sec:dis}. Finally, we also discuss on how to extend the point-particle action to more general couplings in App.~\ref{app} and we complete the expression for the full dissipated dipolar power in App.~\ref{appB}.

Note that we use a different notation than~\cite{goldberger_effective_2006}, in that our Planck mass is related to the (bare) Newton constant by $\Mp^2 = {1}/({8 \pi G_N})$---instead of $m_P^2 = {1}/({32 \pi G_N})$---and our metric signature is $(- + + +)$. This will make some factor of 4 appear in the graviton propagator and will induce some sign differences.

\section{Non-Relativistic Scalar-Tensor Theory} \label{sec3}

It is straightforward to extend the formalism of Goldberger and Rothstein to the scalar-tensor action \eqref{totaction}.
We hereby review  the basics of this approach and highlight the novelty represented by the new degree of freedom.

\subsection{Lengthscales in binary systems}

The EFT formalism of Goldberger and Rothstein, as much as the standard post-Newtonian formalism, is based on the expansion in the small parameter $v$, the velocity of the objects forming the binary system. More formally, $v^2$ is the ratio between the size of the objects $r_s$ (the effective cut-off of the theory, which for black holes and neutron stars is well approximated by the Schwartzschild radius) and the size of the orbit $r$,
 \begin{equation}
v^2 \sim \frac{G_N m}{r} \sim \frac{r_s}{r}\, .
\end{equation}
At the same time, $v$ relates  the orbital size $r$ and the period $T$---equivalently, the wavelength $\lambda$ of the emitted gravitational waves,
\begin{equation}
v \sim \frac{r}{\lambda}\, .
\end{equation}

\subsection{Integrating out fluctuating fields}

As explained above, the binary system breaks Lorentz invariance spontaneously. The formalism goes along with this splitting of spacetime into space and time because,  in order to estimate the powers of $v$ that come from different terms in the action and/or from a given Feynman diagram,  we are suggested to split the metric field fluctuation $h_{\mu \nu}$ into a potential part $H_{\mu \nu}$ and a radiative part  $\bar h_{\mu \nu}$, i.e.,
\begin{equation}
g_{\mu \nu} (x)= g^{(0)}_{\mu \nu} (t) + \frac{h_{\mu \nu} (t,\mathbf{ x})}{\Mp}  \;, \qquad h_{\mu \nu} (t,\mathbf{ x})=   H_{\mu \nu} (t,\mathbf{ x})+ \bar h_{\mu \nu}(t,\mathbf{ x}) \;,
\end{equation}
where $g^{(0)}_{\mu \nu} (t)$ is the background metric.
The difference between the potential and radiative parts is in the scaling of their momenta: emitted gravitons  always have the momentum and the frequency of the binary system ${v}/{r}$, while the spatial momentum of a potential graviton is of the order of the inverse separation between the two components. Denoting the four-momentum of the latter with $k^\mu = (k^0, \mathbf{k})$, one has $k^0 \sim {v}/{r}$ and $\mathbf{k} \sim {1}/{r}$. 

The same separation applies to the scalar field, i.e., 
\begin{equation}
\phi (x)= \phi_0 (t) + \varphi (t,\mathbf{ x})\;, \qquad \varphi (t,\mathbf{ x})= \Phi (t,\mathbf{ x})+ \bar{\varphi}(t,\mathbf{ x})\;, 
\end{equation}
where $\phi_0 (t)$ is the homogeneous time-dependent expectation value of the field.
As we consider systems much smaller than the Hubble radius, we can take the background metric to be the Minkowski metric. Moreover, as we are interested in a dark energy scalar field,  its time variation is of  order Hubble and we can thus neglect it. Therefore, from now on we use 
\be
g^{(0)}_{\mu \nu} = \eta_{\mu \nu} \;, \qquad \phi_0 = \text{const.} \;.
\ee
The constant scalar field VEV   can be reabsorbed in the definition of the masses and scalar charges in eq.~\eqref{eq:action} and can be thus set to zero without loss of generality, $\phi_0=0$.

The effective action is obtained as a two-step path integration, first over  the potential gravitons and scalars, respectively $H_{\mu \nu}$ and $\Phi$, and then over the radiation ones, respectively $\bar h_{\mu \nu}$ and $\bar \varphi$. 
Thus, the first step consists in computing the effective action $ S_{\rm eff}[x_A, \bar{h}_{\mu \nu}, \bar{\varphi}]$, defined by
\be
\label{eq:EFTaction1}
\exp\left({i  S_{\rm eff}[x_A, \bar{h}_{\mu \nu}, \bar{\varphi}]} \right) = \int {\cal D} H_{\mu \nu} {\cal D} \Phi \exp \left({i  S[x_A, {h}_{\mu \nu}, {\varphi}] + i S_{{\rm GF}, H} [H_{\mu \nu}, {\bar h}_{\mu \nu}] } \right) \;,
\ee
where $S_{{\rm GF},H}$ is a gauge-fixing term to the so-called de Donder (or harmonic) gauge, which allows to define the  propagator of $H_{\mu \nu}$. Its expression is given by
\begin{equation}
S_{{\rm GF},H} = - \frac{1}{4} \int d^4x \sqrt{-\bar{g}} \, \bar g^{\mu \nu} \, \Gamma^{(H)}_\mu \Gamma^{(H)}_\nu \;, \qquad \Gamma^{(H)}_\mu \equiv D_\alpha H^\alpha_\mu - \frac{1}{2}D_\mu H^\alpha_\alpha \;,
\label{eq:GFterm}
\end{equation}
where $\bar g_{\mu \nu} \equiv \eta_{\mu \nu} + \bar{h}_{\mu \nu}/\Mp$ is the  background metric for $H_{\mu \nu}$ and
$D_\mu$ is the covariant derivative compatible with it. 
Here we do not consider Faddeev-Popov ghosts because they appear only in loops and we will only compute tree-level diagrams. Indeed, as discussed below loop contributions can be shown to be suppressed with respect to tree level diagrams by the (huge) total angular momentum $L$ of the system~\cite{goldberger_effective_2006}.

The action obtained by this procedure contains the mechanical two-body Lagrangian of the system and the coupling to radiation gravitons. 
For $\bar{h}_{\mu \nu} =0$ and $ \bar{\varphi}=0$ the two body dynamics is conservative and,  to  leading and next to leading order in  $v$  the Lagrangian  reduces   to the Newtonian  and   EIH Lagrangians respectively, extended by the suitable corrections  coming from the scalar fifth force. We compute these in Sec.~\ref{sec:feynman} and \ref{sec:EIH_Lagrangian}.

The second integration, i.e.~over $\bar{h}_{\mu \nu}$ and $\bar{\varphi}$, gives
\be
\label{eq:finalSeff}
\exp \left(i \hat S_{\rm eff}[x_A] \right) = \int \mathcal{D} \bar{h}_{\mu \nu} \mathcal{D} \bar{\varphi} \exp \left( i  S_{\rm eff}[x_A, \bar{h}_{\mu \nu}, \bar{\varphi}] + i S_{{\rm GF}, \bar{h}}  [{\bar h}_{\mu \nu}] \right)  \;,
\ee
where $S_{{\rm GF},\bar h}$ is the gauge-fixing term for $\bar h_{\mu \nu}$, defined as
\begin{equation}
S_{{\rm GF},\bar h} = - \frac{1}{4} \int d^4x  \, \eta^{\mu \nu} \, \Gamma^{(\bar h)}_\mu \Gamma^{(\bar h)} _\nu \;, \qquad \Gamma^{(\bar h)}_\mu \equiv \partial_\alpha \bar h^\alpha_\mu - \frac{1}{2} \partial_\mu \bar h^\alpha_\alpha \;.
\label{eq:GFterm2}
\end{equation}
We have denoted with a hat the final effective action after the metric and scalar fields have been totally integrated out. As we review in Sec.~\ref{sec:dissipative_dynamics},  $\hat S_{\rm eff}[x_A]$ (more precisely, its imaginary part) contains information about the radiated power into gravitational and scalar waves.

\subsection{Propagators and power counting}
\label{sec3.3}

The fields propagators of the gravitational sector can be obtained from the quadratic  action,
\begin{equation}
S^{(2)} = - \frac{1}{8} \int d^4x \left[ -\frac{1}{2}(\partial_\mu h^\alpha_\alpha)^2 + (\partial_\mu h_{\nu \rho})^2 \right]  - \frac12  \int d^4x ( \partial_\mu \varphi )^2 \;,
\end{equation}
where repeated indices are contracted with the Minkowski metric.
In Fourier space it becomes
\begin{equation}
S^{(2)} = - \frac{1}{2} \int \frac{d^4 k}{(2 \pi)^4} k^2 h_{\mu \nu}(k) T^{\mu \nu ; \alpha \beta} h_{\alpha \beta}(-k)  - \frac{1}{2} \int \frac{d^4 k}{(2 \pi)^4} k^2 \varphi(k) \varphi(-k) \;,
\end{equation}
with $k^2 \equiv k_\mu k^\mu$ and $T^{\mu \nu ; \alpha \beta} = \frac{1}{8} \left( \eta^{\mu \alpha} \eta^{\nu \beta} + \eta^{\mu \beta} \eta^{\nu \alpha} - \eta^{\mu \nu} \eta^{\alpha \beta} \right)$.

Let us start discussing the propagators, defined as the inverse quadratic operator. For the metric we have to find the inverse operator of $T^{\mu \nu ; \alpha \beta}$, which is defined as $P_{\mu \nu ; \rho \sigma} T^{\rho \sigma ; \alpha \beta} = I_{\mu \nu}^{\alpha \beta}$ where $I_{\mu \nu}^{\alpha \beta} = \frac{1}{2} (\delta_\mu^\alpha \delta_\nu^\beta + \delta_\mu^\beta \delta_\nu^\alpha)$ is the identity on symmetric two-index tensors. It is straightforward to find 
\be
P_{\mu \nu ; \alpha \beta} = 2 \left( \eta_{\mu \alpha} \eta_{\nu \beta} + \eta_{\mu \beta} \eta_{\nu \alpha} - \eta_{\mu \nu} \eta_{\alpha \beta} \right) \;, 
\ee
where the factor 4 difference with \cite{goldberger_effective_2006} is due to the different normalization of the Planck mass. 
The propagator for the $h$ field is thus given by 
\begin{equation}
\left\langle T h_{\mu \nu}(x) h_{\alpha \beta}(x') \right\rangle = D_F(x-x') P_{\mu \nu ; \alpha \beta} \;,
\label{eq:propagator}
\end{equation}
where $T$ denotes time ordering and the Feynman propagator $D_F(x-x')$ is given by 
\begin{equation}
D_F(x-x') = \int \frac{d^4 k}{(2 \pi)^4} \frac{-i}{k^2-i\epsilon} e^{-ik(x-x')} \;.
\label{eq:Feyprop}
\end{equation}
The term  $i\epsilon$ is the usual prescription for the contour integral.

\begin{figure}
    \centering
    \subfloat[$\bar{h}_{\mu\nu}$]{  \label{fig:propagartorsa}
		\begin{tikzpicture}
			\begin{feynman}
				\vertex (i1);
				\vertex [right=of i1] (a);
				
				\diagram*{
				i1 -- [photon] (a),
				};
				
			\end{feynman}
		
		\end{tikzpicture}
	} \hspace{1em}
	\subfloat[$H_{\mu\nu}$]{  \label{fig:propagartorsb}
		\begin{tikzpicture}
			\begin{feynman}
				\vertex (i1);
				\vertex [right=of i1] (a);
				
				\diagram*{
				i1 -- [scalar] (a),
				};
				
			\end{feynman}
		
		\end{tikzpicture}
	} \hspace{1em}
	\subfloat[$\bar{\phi}$]{  \label{fig:propagartorsc}
		\begin{tikzpicture}
			\begin{feynman}
				\vertex (i1);
				\vertex [right=of i1] (a);
				
				\diagram*{
				i1 -- [gluon] (a),
				};
				
			\end{feynman}
		
		\end{tikzpicture}
	} \hspace{1em}
	\subfloat[$\Phi$]{  \label{fig:propagartorsd}
		\begin{tikzpicture}
			\begin{feynman}
				\vertex (i1);
				\vertex [right=of i1] (a);
				
				\diagram*{
				i1 -- [ghost] (a),
				};
				
			\end{feynman}
		
		\end{tikzpicture}
	} \hspace{1em}
	
    \caption{Representation of the different fields propagators in Feynman diagrams.}
    \label{fig:propagartors}
\end{figure}
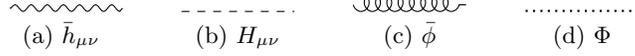
At this point it is useful to make a distinction between the propagator of $\bar{h}_{\mu \nu}$ and that of $H_{\mu \nu}$. While for $\bar{h}_{\mu \nu}$ we must use the relativistic propagator  given in eq.~\eqref{eq:propagator}, for $H_{\mu \nu}$ we can take advantage of the fact that its time and space derivatives scale differently with the velocity $v$, $k^0 \sim {v}/{r} \ll |\mathbf{k} | \sim {1}/{r}$. With the partial (only spatial) Fourier decomposition 
\begin{equation}
H_{\mu \nu}(t,\mathbf{x}) = \int \frac{d^3 k}{(2 \pi)^3} H_{\mathbf{k} \, \mu \nu}(t) e^{i \mathbf{k} \cdot \mathbf{x}} \;,
\label{eq:potential_graviton}
\end{equation}
the $v$ power counting becomes more transparent, as we have $\partial_0 H_{\mathbf{k} \, \mu \nu} \sim v \, \mathbf{k}  H_{\mathbf{k}  \, \mu \nu} $.
Therefore, using the expansion 
\be
\frac{-i}{k^2-i\epsilon} = - \frac{i}{\mathbf{k}^2} \left(1 + \frac{k_0^2}{\mathbf{k}^2} + \dots \right) \;,
\ee
(on the right-hand side we have gotten rid of the $i\epsilon$ prescription, which is irrelevant  for off-shell gravitons), one finds the propagator as the lowest-order term in this expansion,
\begin{equation}
\langle T H_{ \mathbf{k}\, \mu \nu}(t) H_{ \mathbf{q}\, \alpha \beta}(t') \rangle = - (2 \pi)^3 \frac{i}{\mathbf{k}^2} \delta^{(3)}(\mathbf{k} + \mathbf{q}) \delta(t-t') P_{\mu \nu ; \alpha \beta}
\label{eq:MixedFourierPropagator} \;.
\end{equation}
Figures~\ref{fig:propagartorsa} and \ref{fig:propagartorsb} illustrate how  the propagators of $\bar h_{\mu \nu}$ and $H_{\mu \nu}$ are represented in Feynman diagrams.

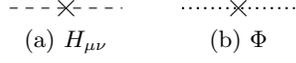
\begin{figure}
    \centering
    \subfloat[$H_{\mu\nu}$]{
		\begin{tikzpicture}
			\begin{feynman}
				\vertex (i1);
				\vertex [right=of i1] (a);
				
				\diagram*{
				i1 -- [scalar, insertion=0.5] (a),
				};
				
			\end{feynman}
		
		\end{tikzpicture}
	} \hspace{1em}
	\subfloat[$\Phi$]{
		\begin{tikzpicture}
			\begin{feynman}
				\vertex (i1);
				\vertex [right=of i1] (a);
				
				\diagram*{
				i1 -- [ghost, insertion=0.5] (a),
				};
				
			\end{feynman}
		
		\end{tikzpicture}
	} \hspace{1em}

    \caption{Representation of the first velocity correction to the potential propagators.}
    \label{fig:insertion}
\end{figure}

The first correction to the propagator of $H_{\mu \nu}$ is supressed by $v^2$ and reads
\begin{equation}
\langle T H_{\mathbf{k} \, \mu \nu}(t) H_{\mathbf{q} \, \alpha \beta}(t') \rangle_{v^2} = - (2 \pi)^3 \frac{i}{\mathbf{k}^4} \delta^{(3)}(\mathbf{k} + \mathbf{q}) \frac{d^2}{dt dt'}\delta(t-t') P_{\mu \nu ; \alpha \beta} \;,
\label{eq:modH}
\end{equation}
which is represented as an insertion on the propagator, as illustrated in Fig.~\ref{fig:insertion}a.

For the treatment of the scalar field dynamics the procedure is analogous---and without the complication of the indices. We define the partial Fourier transform of $\Phi$ by 
\be
\Phi(t,\mathbf{x}) = \int \frac{d^3 k}{(2 \pi)^3} \Phi_{\mathbf{k} }(t) e^{i \mathbf{k} \cdot \mathbf{x}} \;,
\ee
and for the propagators we obtain
\begin{align}
\langle T \bar{\varphi} (x) \bar{\varphi}(x') \rangle &= D_F(x-x') \; ,  \label{eq:propagatorscalar}\\
\langle T \Phi_\mathbf{k} (t) \Phi_\mathbf{q} (t') \rangle &= - (2 \pi)^3 \frac{i}{\mathbf{k}^2} \delta^{(3)}(\mathbf{k} + \mathbf{q}) \delta(t-t') \;. \label{eq:scalarpot}
\end{align}
We display their Feynman diagram representation in Figs.~\ref{fig:propagartors}c and \ref{fig:propagartors}d.
The first correction to the propagator of $\Phi_\mathbf{k}$ reads
\be
\langle T \Phi_\mathbf{k} (t) \Phi_\mathbf{q} (t') \rangle_{v^2} = - (2 \pi)^3 \frac{i}{\mathbf{k}^4} \delta^{(3)}(\mathbf{k} + \mathbf{q}) \frac{d^2}{dt dt'}\delta(t-t')  \;,
\label{eq:modPhi}
\ee
and its representation is illustrated in Fig.~\ref{fig:insertion}b.

In order to organize systematically the Feynman diagrams in powers of $v$ and estimate their contributions to the effective action
we need to find the power counting rules of the theory. The power counting rules for the radiating and potential fields can be extracted from their propagators. For a radiation graviton, the propagator in eq.~\eqref{eq:propagator} is given by eq.~\eqref{eq:Feyprop}. Using  $k \sim {v}/{r}$, this scales as $(v/r)^2$, which gives $\bar{h}_{\mu \nu}(x) \sim {v}/{r}$. The same reasoning applies to $\bar \varphi$, which gives $\bar{\varphi}(x) \sim {v}/{r}$.

For the potential graviton we can use the expression of its propagator, eq.~\eqref{eq:MixedFourierPropagator}. Using that the delta function scales as the inverse of its argument, we obtain that the right-hand side of this equation scales as $k^0 / \mathbf{k}^5$, which using that $k^0 \sim {v}/{r}$ and  $\mathbf{k} \sim {1}/{r}$, gives $H_{\mathbf{k} \, \mu \nu}(t) \sim r^2 \sqrt{v}$. The scaling of the scalar can be found in a similar way and gives $\Phi_\mathbf{k} \sim r^2 \sqrt{v}$.

One last subtlety that we need to address to determine the correct power counting of the theory is the presence of the large parameter  ${m}/{\Mp}$ in the point-particle actions, see eq.~\eqref{eq:action}. As discussed in \cite{goldberger_effective_2006}, this can be resolved  by treating the lowest-order diagrams non-perturbatively, while the next-order diagrams are down by powers of $v$ compared to the leading  ones. 
 In order to make this fact more explicit one 
  introduces the orbital angular momentum associated to the point-particle worldlines,  
\be
L \equiv m v r\;, 
\ee
and  uses the virial relation $v^2 \sim m/(\Mp^2 r)$ to eliminate $m$ and ${m}/{\Mp}$ from the power-counting rules, replacing them by the appropriate combinations of $L$, $v$ and $r$. For instance, one obtains $m/\Mp \sim \sqrt{L v}$.
For the diagrams describing the interactions between the two bodies, the leading  operators scale as $L v^0$ and  must be treated non-perturbatively in accordance 
with the fact that particle worldlines are background non-dynamical fields. Indeed, 
since they
 represent infinitely heavy fields that have been integrated out, they have no associated propagators.

 The large parameter $L$ can be also used to count loop diagrams. Loops that are closed by the particle worldlines are not quantum loops but give tree-level contributions.
 Diagrams with  actual quantum loops (i.e.~not involving particle worldlines) are suppressed by powers of $\hbar/L \ll 1$ and should therefore be discarded.

\subsection{Vertices}

We can now turn to compute the vertices of the action \eqref{eq:action}. They are of two kinds: the ones generated by the Einstein-Hilbert and scalar field kinetic terms, i.e.~$S_{\rm grav}$, and the ones generated by the point-particle action $S_{{\rm pp}}$.

Let us first discuss the first kind. We will focus on vertices that are needed for our calculations, i.e.~cubic vertices containing only potential fields $H_{\mu \nu}$ and $\Phi$ and vertices that are linear in the radiation fields $\bar h_{\mu \nu}$ and $\bar \varphi$. The Einstein-Hilbert term, together with the gauge-fixing term \eqref{eq:GFterm}, generates a $H^3$ and a $\bar{h}H^2$  vertex. These have been computed in \cite{goldberger_effective_2006} (see eqs.~(37) and (45) of that reference) and due to their complexity we do not display them here. The relevant part of the scalar field action  is
\begin{equation}
S_{h\phi^2} = - \frac{1}{2 \Mp} \int d^4x \left( \frac{1}{2} h^\alpha_\alpha \eta^{\mu \nu} - h^{\mu \nu} \right) \partial_\mu \varphi \partial_\nu \varphi \;,
\label{eq:hPhi2Vertex}
\end{equation}
where $h_{\mu \nu} (t, \mathbf{x})= \bar{h}_{\mu \nu}(t, \mathbf{x}) + \int \frac{d^3k}{(2\pi)^3} H_{\mathbf{k}\, \mu \nu}(t) e^{i\mathbf{k} \cdot \mathbf{x}}$  and $ \varphi (t, \mathbf{x}) = \bar{\varphi} (t, \mathbf{x}) + \int \frac{d^3k}{(2\pi)^3} \Phi_\mathbf{k}(t) e^{i\mathbf{k} \cdot \mathbf{x}}$.
This
generates a $H \Phi^2$, a $\bar{h} \Phi^2$ and a $H \bar{\varphi} \Phi$ vertex.

Turning to the point-particle action $S_{ {\rm pp}}$, this can be rewritten, using an affine parameter $\lambda$, as
\begin{equation}
S_{{\rm pp}} = \sum_A m_A \int d \lambda \sqrt{- g_{\mu \nu} \frac{dx_A^\mu}{d\lambda} \frac{dx_A^\nu}{d\lambda} } 
\left[-1 + \aaa_A \frac{\varphi}{\Mp} + \bbb_A \left( \frac{\varphi}{\Mp } \right)^2 \right] \;.
\end{equation}  
We can use reparametrization invariance to set $\lambda$ equal to the local time $t$ of the observer. Using the notation $v_A^\mu(t) =  (1, \mathbf{v}_A)$ and $v_A = |\mathbf{v}_A|$, we arrive at the following expression for $S_{{\rm pp}}$, 
\begin{align}
\begin{split}
S_{{\rm pp}} = \sum_A m_A \int dt & \sqrt{1 - v_A^2 - \frac{h_{\mu \nu}}{\Mp} v_A^\mu v_A^\nu} \left[-1 + \aaa_A \frac{\varphi}{\Mp} + \bbb_A \left( \frac{\varphi}{\Mp} \right)^2 \right] \\
=  \sum_A m_A \int dt& \left[-1 + \frac{v_A^2}{2} + \frac{v_A^4}{8} + O(v^6) \right. \\
& + \frac{h_{00}}{2 \Mp} + \frac{h_{0i}}{\Mp} v_A^i + \frac{h_{ij}}{2 \Mp} v_A^i v_A^j + \frac{h_{00}}{4 \Mp} v_A^2 + O(h v^3) \\
&  + \aaa_A \frac{\varphi}{\Mp} - \aaa_A \frac{\varphi v_A^2}{2 \Mp} + O(\varphi v^3) \\
&  + \left. \frac{h_{00}^2}{8 \Mp^2} + \bbb_A \frac{\varphi^2}{\Mp^2} - \aaa_A \frac{\varphi h_{00}}{2 \Mp^2} + O(h^2v, \varphi^2v, h\varphi v) \right] \;,
\label{eq:pp_Action_expanded}
\end{split}
\end{align}
where in the second equality we have expanded the Lagrangian up to order $v^5$.

To get the vertices from the action, one should multiply by $i$ and specify $h_{\mu \nu}$ and $\varphi$ to radiation  or  potential fields. In order to facilitate the power-counting needed to evaluate the order of a Feynman diagram in the expansion in $v$, Table \ref{table:power_counting} sums up the power-counting of the different vertices obtained in this section, which will be needed in the following.

\begin{table}
\center
\begin{tabular}{|c|c|}
\hline
Operator & PCR  \\
\hline
$\displaystyle{m_A \int dt v_A^2}$ & $L$ \\
\hline
$\displaystyle{\aaa_A \frac{m_A}{\Mp} \int dt \varphi, \quad \frac{m_A}{2 \Mp}\int dt h_{00}}$ & $\sqrt{L}$ \\
\hline
$\displaystyle{\frac{m_A}{\Mp} \int dt v_A^i h_{0i}}$ & $\sqrt{L} v$ \\
\hline
$\displaystyle{m_A \int dt \frac{v_A^4}{8}}$ & $L v^2$ \\
\hline
$\displaystyle{\frac{m_A}{4 \Mp} \int dt h_{00}v_A^2, \quad \frac{m_A}{2 \Mp} \int dt h_{ij}v_A^i v_A^j, \quad -\aaa_A \frac{m_A}{2 \Mp} \int dt \varphi v_A^2}$ & $\sqrt{L}v^2$ \\
\hline
$\displaystyle{\frac{m_A}{8 \Mp^2} \int dt h_{00}^2, \quad \bbb_A\frac{m_A}{\Mp^2} \int dt \varphi^2, \quad -\aaa_A\frac{m_A}{2 \Mp^2} \int dt \varphi h_{00}}$ & $v^2$ \\
\hline
$h^3$ from $\displaystyle{\frac{\Mp^2}{2} \int d^4x \sqrt{-g}R}$ (see ref \cite{goldberger_effective_2006} for the explicit expression), $\displaystyle{h \varphi^2}$ from eq.~\eqref{eq:hPhi2Vertex} & $\displaystyle{\frac{v^2}{\sqrt{L}}}$ \\
\hline

\end{tabular}
\caption{Power-counting rules for the vertices obtained by expanding the action, given here for a potential graviton and scalar field. Multiply by $\sqrt{v}$ if needed to replace by a radiation graviton or scalar field.}
\label{table:power_counting}
\end{table}

\subsection{Feynman rules}
\label{sec:feynman}

In this subsection we will give the Feynman rules and, for pedagogical purposes, calculate some simple diagrams. The full set of Feynman rules can be summed up as follows:

\begin{itemize}

\item At a given order in $v$, draw all the diagrams that remain connected when removing the worldlines of the particles, discarding quantum (i.e.~not involving particle worldlines) loop diagrams.

\item For each vertex, multiply the corresponding expression in the Einstein-Hilbert action, in eqs.~\eqref{eq:hPhi2Vertex} and \eqref{eq:pp_Action_expanded} by $i$ and specify $h_{\mu \nu}$ to $\bar{h}_{\mu \nu}$ or $H_{\mu \nu}$ and $\varphi$ to $\bar{\varphi}$ or $\Phi$, taking into account the associated corresponding power counting rules.

\item Contract all the internal graviton or scalar lines. This gives a combinatorial factor corresponding to the number of Wick contractions. An internal potential graviton line corresponds to multiplying by eq.~\eqref{eq:MixedFourierPropagator}, while an internal radiation one corresponds to a multiplication by eq.~\eqref{eq:propagator}. An internal potential scalar line corresponds to multiplying by eq.~\eqref{eq:scalarpot}, while an internal radiation one corresponds to a multiplication by eq.~\eqref{eq:propagatorscalar}.

\item The combinatorial factor can be obtained from the explicit definition of the effective action, $e^{iS{\rm eff}} = \int {\cal D} h {\cal D} \varphi e^{i(S_0+S_{\rm int})}$ where $S_0$ is the quadratic action and $S_{\rm int}$ contains the vertices.
The rule of thumb is the following: divide by the symmetry factor  of the diagram (coming from the fact that for $n$ vertices, the $1/n!$ of the exponential is not always compensated by the rearrangement of the $(\mathrm{vertex})^n$ term if there are identical vertices) and then multiply by the number of different Wick contractions giving the diagram.

\end{itemize}

Note that if we focus only on the integration over the potential fields so as to obtain $ S_\mathrm{eff}[x_A, \bar{h}, \bar{\varphi}]$, potential gravitons and scalars, respectively $H_{\mu \nu}$ and $\Phi$, can only enter Feynman diagrams as internal lines, while radiation gravitons and scalars, respectively $\bar{h}_{\mu \nu}$ and $\bar{\varphi}$, can only enter Feynman diagrams as external lines, i.e.~they cannot be used as propagators.

Let's now look at the calculation of simple diagrams. The simplest is given in Fig.~\ref{fig:Newt_pot_A}, and represents the Newtonian potential. Using the Feynman rules, we have
\begin{align}
\begin{split}
\left. iS_\mathrm{eff}\right|_{\ref{fig:Newt_pot_A}} &= \left[i \frac{m_1}{2 \Mp} \int dt_1 \int \frac{d^3k_1}{(2\pi)^3} e^{i \mathbf{k}_1 \cdot \mathbf{x}_1(t_1)} \right] \left[i \frac{m_2}{2\Mp} \int dt_2 \int \frac{d^3k_2}{(2\pi)^3} e^{i \mathbf{k}_2 \cdot \mathbf{x}_2(t_2)} \right] \\
& \times \left\langle T H_{00}(t_1, \mathbf{k}_1) H_{00}(t_2, \mathbf{k}_2) \right\rangle \\
&= i P_{00;00} \frac{m_1m_2}{4\Mp^2} \int dt \int \frac{d^3k}{(2\pi)^3} \frac{e^{i \mathbf{k} \cdot (\mathbf{x}_1(t)-\mathbf{x}_2(t) )}}{k^2} \\
&= i \int dt \frac{G_N m_1 m_2}{r(t)} \;,
\end{split}
\end{align}
where for the last equality we used that
\be
\int \frac{d^3 k}{(2 \pi)^3} \frac{e^{-i \mathbf{k} \cdot \mathbf{x}}}{\mathbf{k}^2} = \frac{1}{4 \pi |\mathbf{x}|} \;,
\ee
and  we have defined $\mathbf{r} \equiv \mathbf{x}_1 - \mathbf{x}_2$ and $r \equiv |\mathbf{r}|$.
An analogous calculation can be done for the scalar interaction given by Fig.~\ref{fig:Newt_pot_b}. This yields
\be
\left. iS_\mathrm{eff}\right|_{\ref{fig:Newt_pot_b}} =  i \int dt \frac{2 G_N \aaa_1 \aaa_2 m_1 m_2}{r(t) } \;,
\ee
so that the effective gravitational Newton constant between two objects $A$ and $B$ reads
\be
\tilde G_{AB} \equiv G_N (1+2 \aaa_A \aaa_B ) \;.
\label{eq:Gab}
\ee

A second example with non-trivial symmetry factor is given by Fig.~\ref{subfig:EIH_f} below, i.e.
\be
\begin{split}
\left. iS_\mathrm{eff}\right|_{\ref{subfig:EIH_f}} &= \frac{1}{2!} \left[i \frac{m_1}{8\Mp^2} \int dt_1 \int \frac{d^3k_1}{(2\pi)^3} \frac{d^3k'_1}{(2\pi)^3} e^{i (\mathbf{k}_1 + \mathbf{k}'_1) \cdot \mathbf{x}_1(t_1)} \right] \left[i \frac{m_2}{2\Mp} \int dt_2 \int \frac{d^3k_2}{(2\pi)^3} e^{i \mathbf{k}_2 \cdot \mathbf{x}_2(t_2)} \right] \\
& \times \left[i \frac{m_2}{2\Mp} \int dt'_2 \int \frac{d^3k'_2}{(2\pi)^3} e^{i \mathbf{k}'_2 \cdot \mathbf{x}_2(t'_2)} \right] \left\langle T H_{00}(t_1, \mathbf{k}_1) H_{00}(t_1, \mathbf{k}'_1) H_{00}(t_2, \mathbf{k}_2) H_{00}(t'_2, \mathbf{k}'_2) \right\rangle \\
&= \frac{i m_1 m_2^2 P_{00;00}^2}{2^5\Mp^4} \int dt \left(\int \frac{d^3k}{(2\pi)^3}  \frac{e^{i \mathbf{k} \cdot (\mathbf{x}_1(t)-\mathbf{x}_2(t))}}{k^2} \right)^2 \\
&= i \int dt \frac{m_1 m_2^2 G_N^2}{2 r^2} \;.
\end{split}
\ee
Now we can move to the complete calculation of the effective action.

\section{Renormalization of  masses and charges}\label{sec2.2}

The action \eqref{eq:action} contains body-dependent scalar charges $\aaa_A$ and $\bbb_A$. As mentioned in the introduction, even if we assume the validity of the weak equivalence principle---the universality of free falling for test particles---in a scalar-tensor theory such a universality is inevitably spoiled by the gravitational self-energy of massive bodies. It is instructive to see how this happens in the adopted formalism. To this end, in this section, we derive the dependence of the scalar charges on the gravitational self-energy, after computing how the masses of the objects get similarly renormalized. 

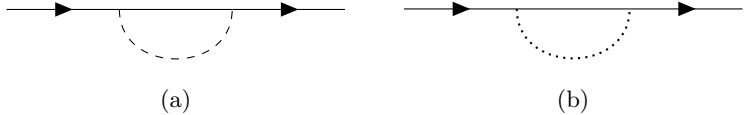
\begin{figure}
	\centering
	\subfloat[]{
		\begin{tikzpicture}
			\begin{feynman}
				\vertex (i1);
				\vertex [right=of i1] (a);
				\vertex [right=of a] (b);
				\vertex [right=of b] (f1);
				
				\diagram*{
				i1 -- [fermion] (a) -- (b) -- [fermion] (f1),
				(a) -- [scalar, half right] (b)
				};
				
			\end{feynman}
		
		\end{tikzpicture}
	} \hspace{1em}
	\subfloat[]{
		\begin{tikzpicture}
			\begin{feynman}
				\vertex (i1);
				\vertex [right=of i1] (a);
				\vertex [right=of a] (b);
				\vertex [right=of b] (f1);
				
				\diagram*{
				i1 -- [fermion] (a) -- (b) -- [fermion] (f1),
				(a) -- [ghost, half right] (b)
				};
				
			\end{feynman}
		
		\end{tikzpicture}
	} \hspace{1em}
	
\caption{Diagrams contributing to the mass renormalization.}
\label{fig:mass_renormalization}
\end{figure}

Let us consider the point-particle action \eqref{eq:action} in the static case (i.e.~for $v_A=0$), focussing on a single body.
To simplify the notation, we will omit the index $A$ in this part of the discussion. The action then reads
\be
\label{eq:m}
- m \int d  \tau \left(1 - \alpha \frac{\varphi}{\Mp} - \beta \left( \frac{\varphi}{\Mp} \right)^2 \right)    \;.
\ee
We want to show that the mass $m$ gets renormalized by the contribution of the self-energy of the  object. 
In particular, at lowest order the two diagrams of Fig.~\ref{fig:mass_renormalization} contribute to this action. In the previous derivation, we ignored such diagrams because they are scale-less power-law divergent and, as such, they vanish in dimensional regularization. However, here we will be concerned about the physical significance of such self-energy diagrams, so we choose instead a hard cutoff $\Lambda$ in the momentum integrals, corresponding approximatively to choosing an object of size $r_s \sim 1/\Lambda $. This regularization preserves rotational symmetry. The fact that, on the other hand, it breaks boosts does not concern us too much here as we are considering  objects at rest.

Starting from a bare mass $m_{\text{bare}}$, by including these diagrams we obtain a dressed mass $m (\Lambda)$. In other words, the term in~\eqref{eq:m} that is of lowest order in the fields gets renormalized as
\be
\label{eq:dressedm}
- m_{\text{bare}} \int dt  \quad \to \quad - m (\Lambda) \int dt  \;,
\ee
where 
\begin{equation}
\label{eq:dm}
m (\Lambda) \equiv m_{\text{bare}} + \delta m(\Lambda) \;, \qquad \delta m(\Lambda)  = -  2 \pi  \tilde{G} m_{{\rm bare}}^2 \,  \int^\Lambda \frac{d^3k}{(2\pi)^3} \frac{1}{k^2}  \;, 
\end{equation}
where we have used the Planck mass definition and we have defined
\be
\tilde{G} \equiv G_N(1+2\aaa_{\rm bare}^2) \;.
\ee 
This (negative) quantity coincides with the gravitational energy of the object, given by the usual expression
\begin{equation}
E = - \frac{\tilde{G}}{2} \int d^3x d^3y \frac{\rho(\mathbf{x}) \rho(\mathbf{y})}{|\mathbf{x} - \mathbf{y}|} \;,
\end{equation}
where $\rho$ is the mass density of the object. Indeed, 
replacing the point-particle density by a regularized version of a delta function, the energy density can be expressed as
\be
\rho(\mathbf{x} )= m_{\rm bare} \int^\Lambda \frac{d^3 k}{(2 \pi)^3} {e^{i \mathbf{k} \cdot \mathbf{x} }} \;,
\ee
and comparing this expression with the second equality in eq.~\eqref{eq:dm}, one obtains
\be
\label{eq:genergy}
E(\Lambda) \equiv   \delta m(\Lambda) \;.
\ee

\begin{figure}
	\centering
	\subfloat[]{
		\begin{tikzpicture}
			\begin{feynman}
				\vertex (i1);
				\vertex [right=of i1] (a);
				\vertex [right=of a] (b);
				\vertex [right=of b] (f1);
				\vertex [below=of a] (c);
				
				\diagram*{
				i1 -- [fermion] (a) -- (b) -- [fermion] (f1),
				(a) -- [scalar] (c),
				(a) -- [scalar, half right] (b)
				};
				
			\end{feynman}
		
		\end{tikzpicture}
	} \hspace{1em}
	\subfloat[]{
		\begin{tikzpicture}
			\begin{feynman}
				\vertex (i1);
				\vertex [right=of i1] (a);
				\vertex [right=of a] (b);
				\vertex [right=of b] (f1);
				\vertex [below=of a] (c);
				
				\diagram*{
				i1 -- [fermion] (a) -- (b) -- [fermion] (f1),
				(a) -- [scalar] (c),
				(a) -- [ghost, half right] (b)
				};
				
			\end{feynman}
		
		\end{tikzpicture}
	} \hspace{1em}
	\subfloat[]{
		\begin{tikzpicture}
			\begin{feynman}
				\vertex (i1);
				\vertex [right=2em of i1] (a);
				\vertex [right=2em of a] (t1);
				\vertex [right=2em of t1] (b);
				\vertex [right=2em of b] (f1);
				\vertex [below=2em of t1] (c);
				\vertex [below=2em of c] (f2);
				
				\diagram*{
				(i1) -- [fermion] (a) -- (t1) -- (b) -- [fermion] (f1),
				(a) -- [scalar] (c),
				(b) -- [scalar] (c),
				(c) -- [scalar] (f2),
				};
				
			\end{feynman}
		
		\end{tikzpicture}
	} \hspace{1em}
	\subfloat[]{
		\begin{tikzpicture}
			\begin{feynman}
				\vertex (i1);
				\vertex [right=2em of i1] (a);
				\vertex [right=2em of a] (t1);
				\vertex [right=2em of t1] (b);
				\vertex [right=2em of b] (f1);
				\vertex [below=2em of t1] (c);
				\vertex [below=2em of c] (f2);
				
				\diagram*{
				(i1) -- [fermion] (a) -- (t1) -- (b) -- [fermion] (f1),
				(a) -- [ghost] (c),
				(b) -- [ghost] (c),
				(c) -- [scalar] (f2),
				};
				
			\end{feynman}
		
		\end{tikzpicture}
	}
	
\caption{Diagrams contributing to the mass renormalization of the vertex $\frac{m}{2 \Mp} \int dt h_{00}$.}
\label{fig:mass_renormalization_higher_order}
\end{figure}
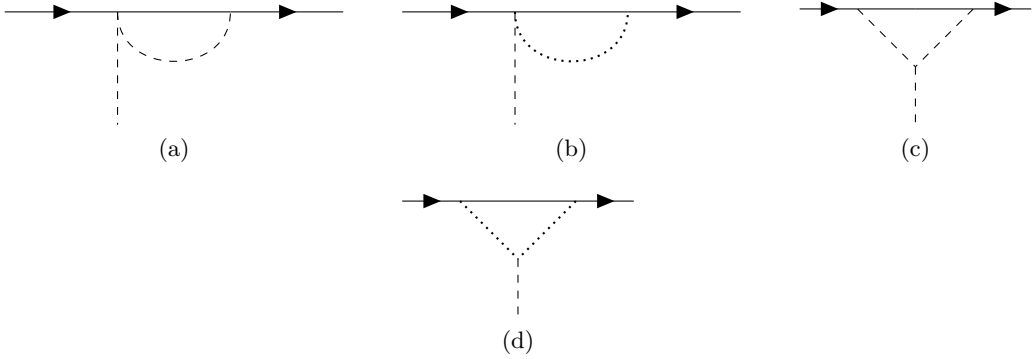

Here we have studied the renormalization of the particle mass appearing in the lowest-order vertex \eqref{eq:m} but, by the equivalence principle, the same mass appears also in 
higher-order operators of the point-particle action \eqref{eq:action}, such as 
\be
\frac{m}{2 \Mp} \int dt \, h_{00}\;.
\ee 
If the use of our hard-cutoff regulator is  consistent, we should get the same result for the renormalization of this vertex. Indeed, we have  checked that this is the case by calculating the Feynman diagrams shown in Fig.~\ref{fig:mass_renormalization_higher_order}. We will not reproduce this lengthy computation here, as it parallels the one for the scalar coupling that we are going to discuss next, with the complication of the spin-2 vertex. The final result is the same mass renormalization as in the lowest-order vertex \eqref{eq:m}, i.e.,  
\be
\frac{m_\text{bare}}{2 \Mp} \int dt \, h_{00} \quad \to \frac{m (\Lambda)}{2 \Mp} \int dt \, h_{00} \;,
\ee 
which shows the consistency of the method
\footnote{In calculating the diagrams of Fig.~\ref{fig:mass_renormalization_higher_order}, there appears also terms proportional to $h_{ij} \delta^{ij}$ in the point-particle action. Of course, for a particle at rest such terms should not appear, as in the proper time the only combination involving $h_{\mu \nu}$ is $h_{\mu \nu} v^\mu v^\nu = h_{00}$. We can trace back the appearance of such artifacts from the fact that our regulator breaks Lorentz invariance.}.

We will now address the renormalization of the scalar charges of the objects $A$, appearing in the operator
\be
\label{eq:dressed}
\aaa_A \frac{m_A}{\Mp} \int dt \, \varphi \;.
\ee
In particular, we will show that even if we assume the weak equivalence principle and start with the same bare scalar  charges for the two objects, $\aaa_{\text{bare},1}=\aaa_{\text{bare},2}=\aaa_\text{bare}$, these get renormalized by the higher-order interactions, 
\be
\aaa_\text{bare} \frac{m_{\text{bare},A}}{\Mp} \int dt \, \varphi \qquad \to \qquad  \aaa_A (\Lambda) \frac{m_{A}(\Lambda)}{\Mp} \int dt \, \varphi \;,
\ee
where
\be
\label{eq:renalpha}
\aaa_A (\Lambda) \equiv \aaa_\text{bare} + \delta \aaa_A (\Lambda) \;.
\ee

To study which diagrams contribute to the charge it is convenient to split the scalar field fluctuation into a potential mode, which we will integrate out, and an external source,  i.e.~$\varphi = \Phi + \varphi_\mathrm{ext}$.
After using such a splitting in the action \eqref{eq:pp_Action_expanded},  the vertices contributing to the renormalization of the operator \eqref{eq:dressed} are
\begin{equation}
-\aaa_\text{bare} m_{\text{bare},A} \int dt \frac{\varphi_\mathrm{ext}H_{00}}{2 \Mp^2} \;, \qquad 2 \bbb_\text{bare} m_{\text{bare},A} \int dt \frac{\varphi_\mathrm{ext} \Phi}{\Mp^2} \;, 
\end{equation}
where for these bare vertices we have assumed the weak equivalence principle, i.e.~that the scalar couplings $\aaa_\text{bare}$ and $\bbb_\text{bare}$ are common to the two objects. 
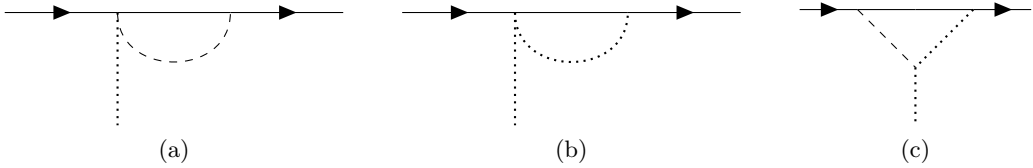
\begin{figure}
	\centering
	\subfloat[]{
		\begin{tikzpicture}
			\begin{feynman}
				\vertex (i1);
				\vertex [right=of i1] (a);
				\vertex [right=of a] (b);
				\vertex [right=of b] (f1);
				\vertex [below=of a] (c);
				
				\diagram*{
				i1 -- [fermion] (a) -- (b) -- [fermion] (f1),
				(a) -- [ghost] (c),
				(a) -- [scalar, half right] (b)
				};
				
			\end{feynman}
		\label{fig:charge_renormalization_a}
		\end{tikzpicture}
	} \hspace{1em}
	\subfloat[]{
		\begin{tikzpicture}
			\begin{feynman}
				\vertex (i1);
				\vertex [right=of i1] (a);
				\vertex [right=of a] (b);
				\vertex [right=of b] (f1);
				\vertex [below=of a] (c);
				
				\diagram*{
				i1 -- [fermion] (a) -- (b) -- [fermion] (f1),
				(a) -- [ghost] (c),
				(a) -- [ghost, half right] (b)
				};
				
			\end{feynman}
		\label{fig:charge_renormalization_b}
		\end{tikzpicture}
	} \hspace{1em}
	\subfloat[]{
		\begin{tikzpicture}
			\begin{feynman}
				\vertex (i1);
				\vertex [right=2em of i1] (a);
				\vertex [right=2em of a] (t1);
				\vertex [right=2em of t1] (b);
				\vertex [right=2em of b] (f1);
				\vertex [below=2em of t1] (c);
				\vertex [below=2em of c] (f2);
				
				\diagram*{
				(i1) -- [fermion] (a) -- (t1) -- (b) -- [fermion] (f1),
				(a) -- [scalar] (c),
				(b) -- [ghost] (c),
				(c) -- [ghost] (f2),
				};
				
			\end{feynman}
		\label{fig:charge_renormalization_c}
		\end{tikzpicture}
	}
	
\caption{Diagrams contributing to the charge renormalization.}
\label{fig:charge_renormalization}
\end{figure}
The  corresponding  diagrams are  shown in Fig.~\ref{fig:charge_renormalization_a} and \ref{fig:charge_renormalization_b}. The correction to the renormalized scalar charge appearing in the vertex $\aaa_A(\Lambda)  \frac{m_A}{\Mp} \int dt \, \varphi_{\rm ext} $
reads
\begin{equation}
\delta \aaa_A (\Lambda)= \left( 2 \aaa  \frac{1-4 \bbb}{1+2 \aaa^2} \right)_{\rm bare} \frac{\delta m_A(\Lambda)}{m_{A}(\Lambda)} \;, 
\label{eq:deltaalpha}
\end{equation}
where for $\delta m_A$ and $m_{A}$ on the right-hand side we have used eqs.~\eqref{eq:dm} and \eqref{eq:genergy}.
A third vertex, represented in Fig.~\ref{fig:charge_renormalization_c}, comes from the scalar field action \eqref{eq:hPhi2Vertex} and reads
\begin{equation}
 - \frac{1}{\Mp} \int d^4x \left( \frac{H^\alpha_\alpha}{2} \eta^{\mu \nu} - H^{\mu \nu} \right) \partial_\mu \varphi_\mathrm{ext} \partial_\nu \Phi \;.
\end{equation}
However, the tensorial factor in the  propagator of the potential graviton modes is non-vanishing only for $\mu=0$ and $\nu=0$, which implies that this diagram vanishes in our static case.

Plugging the renormalized values of the scalar couplings \eqref{eq:renalpha} with eq.~\eqref{eq:deltaalpha} in the expression for the effective Newton constant between two bodies $A$ and $B$ given in eq.~\eqref{eq:Gab} and expanding to leading order in $\delta \aaa$, we obtain
\begin{equation}
\tilde{G}_{AB} \simeq \tilde{G}\left[ 1+ 4\aaa^2 \frac{1 - 4 \bbb}{(1+2 \aaa^2)^2} \left( \frac{E_A}{m_A} + \frac{E_B}{m_B} \right) \right] \;.
\end{equation}
As first realized by Nordtvedt \cite{Nordtvedt:1968qs}, this implies a violation of the strong Equivalence Principle. This expression agrees with the one derived in Ref.~\cite{damour_tensor-multi-scalar_1992}, which shows that the body-dependent gravitational constant $\tilde{G}_{AB}$ is given by 
\begin{equation}
\tilde{G}_{AB} = \tilde{G}\left[ 1+(4 \tilde \beta - \tilde \gamma - 3) \left( \frac{E_A}{m_A} + \frac{E_B}{m_B} \right) \right] \;,
\end{equation}
where $\tilde \beta$ and $\tilde \gamma$ are the parametrized post-Newtonian (PPN) parameters, given  by\footnote{The PPN parameters are given in eqs.~(4.12b) and (4.12c) of Ref.~\cite{damour_tensor-multi-scalar_1992}, where we  the dictionary between our notation and theirs is $\varphi_{\rm here} = \sqrt{2}M_P \varphi_{\rm there}$, $\alpha^a= -\aaa \sqrt{2}$, $\beta_{ab} = (-4\bbb -2\aaa^2) \delta_{ab}$.}
\begin{align}
\tilde \beta &= 1 - 2 \left[ \frac{\aaa ^4+2 \aaa ^2 \bbb }{(1+2 \aaa ^2)^2} \right]_{\rm bare}\;, \\
\tilde \gamma &= 1 - 4 \left[ \frac{\aaa^2}{1+2 \aaa^2} \right]_{\rm bare}\;.
\end{align}
Considerations similar to the one for the renormalization of  $\aaa_A$ can be made for the couplings $\bbb_A$.

Before concluding the section, let us notice that the renormalization of the scalar charges can be also expressed in terms of the so-called ``sensitivity''
of a body to changes in the local value of the effective gravitational constant $G_N$ due to
changes in the scalar field \cite{Eardley1975ApJ}. It is explicitely defined by
\be
\label{eq:sensitivity}
s_A \equiv - \frac{d \ln m_A}{d \ln G_N}  = - \frac{\delta m_A}{m_A}\;,
\ee
where in the last equality we have used eq.~\eqref{eq:dm}. Using eq.~\eqref{eq:deltaalpha}, the sensitivity can be related to the scalar charges by
\be
s_A = -    \left( \frac{1+2 \aaa^2}{1-4 \bbb }\right)_{\rm bare}  \frac{\delta \alpha_A}{2 \aaa_{\rm bare}} \;.
\ee

\section{Conservative dynamics up to $1$PN order}


\label{sec:EIH_Lagrangian}

In Sec.~\ref{sec3} we have shown how to compute the effective action $ S_{\rm eff} [x_A, \bar h_{\mu \nu}, \bar \varphi]$ defined in eq.~\eqref{eq:EFTaction1} by integrating out the potential modes $H_{\mu \nu}$ and $\Phi$. We now focus on the conservative part of this action obtained 
by considering only diagrams without external (and internal) radiation, i.e.~$ S_\mathrm{eff}[x_A, \bar{h}_{\mu \nu}=0, \bar{\varphi}=0]$.

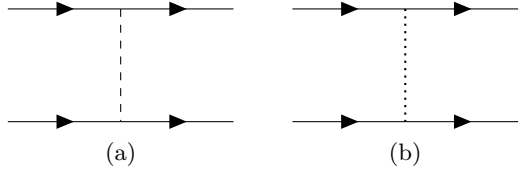
\begin{figure}
	\centering
	\subfloat[]{
		\begin{tikzpicture}
			\begin{feynman}
				\vertex (i1);
				\vertex [right=of i1] (a);
				\vertex [right=of a] (f1);
				\vertex [below=of i1] (i2);
				\vertex [below=of a] (b);
				\vertex [below=of f1] (f2);
				
				\diagram*{
				i1 -- [fermion] (a) -- [fermion] (f1),
				(i2) -- [fermion] (b) -- [fermion] (f2),
				(a) -- [scalar] (b)
				};
			\end{feynman}
			
			\label{fig:Newt_pot_A}
		
		\end{tikzpicture}
	} \hspace{1em}
	\subfloat[]{
		\begin{tikzpicture}
			\begin{feynman}
				\vertex (i1);
				\vertex [right=of i1] (a);
				\vertex [right=of a] (f1);
				\vertex [below=of i1] (i2);
				\vertex [below=of a] (b);
				\vertex [below=of f1] (f2);
				
				\diagram*{
				i1 -- [fermion] (a) -- [fermion] (f1),
				(i2) -- [fermion] (b) -- [fermion] (f2),
				(a) -- [ghost] (b)
				};
			\end{feynman}
		
		\label{fig:Newt_pot_b}
		
		\end{tikzpicture}
	}
	
\caption{Feynman diagrams contributing to the $Lv^0$ potential}
\label{fig:Newt_pot}
\end{figure}
At lowest order in $v$, there are only two diagrams contributing to this action, illustrated in Fig.~\ref{fig:Newt_pot}: respectively one graviton  and one scalar exchange. Therefore, the action to order $Lv^0$  is given by 
\begin{equation}
S_{Lv^0} = \int dt \left[\frac{1}{2} m_1 v_1^2 + \frac{1}{2} m_2 v_2^2 + \frac{\tilde G_{12} m_1 m_2}{r}\right] \;.
\end{equation}
The first two terms correspond to the Newtonian kinetic energy of the particles whereas the last term is the effective gravitational potential with the rescaled Newton constant $\tilde{G}_{AB}$ computed in Sec.~\ref{sec:feynman}, see eq.~\eqref{eq:Gab}.

Let us now compute the first relativistic correction to this result. For GR, the corresponding Lagrangian has been calculated for the first time by Einstein, Infeld and Hoffmann \cite{einstein_gravitational_1938}. It was generalized to   multi-scalar-tensor theories of gravitation by Damour and Esposito-Far\`{e}se  \cite{damour_tensor-multi-scalar_1992} using a post-Newtonian expansion. In the final equation of this section, see eq.~\eqref{eq:LEIH} below, we will recover the result of \cite{damour_tensor-multi-scalar_1992} restricted to a single scalar.

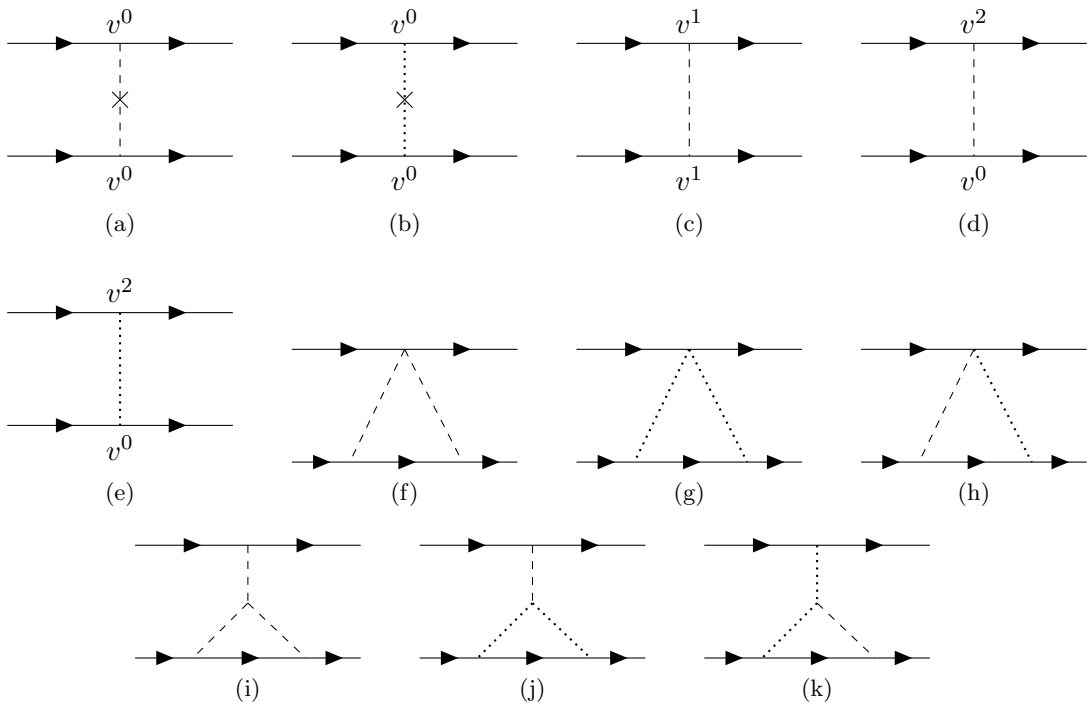
\begin{figure}
	\centering
	\subfloat[]{
		\begin{tikzpicture}
			\begin{feynman}
				\vertex (i1);
				\vertex [right=of i1] (a);
				\vertex [right=of a] (f1);
				\vertex [below=of i1] (i2);
				\vertex [below=of a] (b);
				\vertex [below=of f1] (f2);
				
				\diagram*{
				(i1) -- [fermion] (a) -- [fermion] (f1),
				(i2) -- [fermion] (b) -- [fermion] (f2),
				(a) -- [scalar, insertion=0.5] (b),
				};
				
				\draw[] (a.north) node[above] {$v^0$};
				\draw[] (b.south) node[below] {$v^0$};
			\end{feynman}
		
		\end{tikzpicture}
		\label{subfig:EIH_A}
	} \hspace{1em}
	\subfloat[]{
		\begin{tikzpicture}
			\begin{feynman}
				\vertex (i1);
				\vertex [right=of i1] (a);
				\vertex [right=of a] (f1);
				\vertex [below=of i1] (i2);
				\vertex [below=of a] (b);
				\vertex [below=of f1] (f2);
				
				\diagram*{
				(i1) -- [fermion] (a) -- [fermion] (f1),
				(i2) -- [fermion] (b) -- [fermion] (f2),
				(a) -- [ghost, insertion=0.5] (b),
				};
				
				\draw[] (a.north) node[above] {$v^0$};
				\draw[] (b.south) node[below] {$v^0$};
			\end{feynman}
		
		\end{tikzpicture}
		\label{subfig:EIH_b}
	} \hspace{1em}
	\subfloat[]{
		\begin{tikzpicture}
			\begin{feynman}
				\vertex (i1);
				\vertex [right=of i1] (a);
				\vertex [right=of a] (f1);
				\vertex [below=of i1] (i2);
				\vertex [below=of a] (b);
				\vertex [below=of f1] (f2);
				
				\diagram*{
				(i1) -- [fermion] (a) -- [fermion] (f1),
				(i2) -- [fermion] (b) -- [fermion] (f2),
				(a) -- [scalar] (b),
				};
				
				\draw[] (a.north) node[above] {$v^1$};
				\draw[] (b.south) node[below] {$v^1$};
			\end{feynman}
		
		\end{tikzpicture}
		\label{subfig:EIH_c}
	} \hspace{1em}
	\subfloat[]{
		\begin{tikzpicture}
			\begin{feynman}
				\vertex (i1);
				\vertex [right=of i1] (a);
				\vertex [right=of a] (f1);
				\vertex [below=of i1] (i2);
				\vertex [below=of a] (b);
				\vertex [below=of f1] (f2);
				
				\diagram*{
				(i1) -- [fermion] (a) -- [fermion] (f1),
				(i2) -- [fermion] (b) -- [fermion] (f2),
				(a) -- [scalar] (b),
				};
				
				\draw[] (a.north) node[above] {$v^2$};
				\draw[] (b.south) node[below] {$v^0$};
			\end{feynman}
		
		\end{tikzpicture}
		\label{subfig:EIH_d}
	} \hspace{1em}
	\subfloat[]{
		\begin{tikzpicture}
			\begin{feynman}
				\vertex (i1);
				\vertex [right=of i1] (a);
				\vertex [right=of a] (f1);
				\vertex [below=of i1] (i2);
				\vertex [below=of a] (b);
				\vertex [below=of f1] (f2);
				
				\diagram*{
				(i1) -- [fermion] (a) -- [fermion] (f1),
				(i2) -- [fermion] (b) -- [fermion] (f2),
				(a) -- [ghost] (b),
				};
				
				\draw[] (a.north) node[above] {$v^2$};
				\draw[] (b.south) node[below] {$v^0$};
			\end{feynman}
		
		\end{tikzpicture}
		\label{subfig:EIH_e}
	} \hspace{1em}
	\subfloat[]{
		\begin{tikzpicture}
			\begin{feynman}
				\vertex (i1);
				\vertex [right=of i1] (a);
				\vertex [right=of a] (f1);
				\vertex [below=of i1] (i2);
				\vertex [right=2em of i2] (b);
				\vertex [right=of b] (c);
				\vertex [below=of f1] (f2);
				
				\diagram*{
				(i1) -- [fermion] (a) -- [fermion] (f1),
				(i2) -- [fermion] (b) -- [fermion] (c) -- [fermion] (f2),
				(a) -- [scalar] {(b), (c)},
				};
				
			\end{feynman}
		
		\end{tikzpicture}
		\label{subfig:EIH_f}
	} \hspace{1em}
	\subfloat[]{
		\begin{tikzpicture}
			\begin{feynman}
				\vertex (i1);
				\vertex [right=of i1] (a);
				\vertex [right=of a] (f1);
				\vertex [below=of i1] (i2);
				\vertex [right=2em of i2] (b);
				\vertex [right=of b] (c);
				\vertex [below=of f1] (f2);
				
				\diagram*{
				(i1) -- [fermion] (a) -- [fermion] (f1),
				(i2) -- [fermion] (b) -- [fermion] (c) -- [fermion] (f2),
				(a) -- [ghost] {(b), (c)},
				};
				
			\end{feynman}
		
		\end{tikzpicture}
		\label{subfig:EIH_g}
	} \hspace{1em}
	\subfloat[]{
		\begin{tikzpicture}
			\begin{feynman}
				\vertex (i1);
				\vertex [right=of i1] (a);
				\vertex [right=of a] (f1);
				\vertex [below=of i1] (i2);
				\vertex [right=2em of i2] (b);
				\vertex [right=of b] (c);
				\vertex [below=of f1] (f2);
				
				\diagram*{
				(i1) -- [fermion] (a) -- [fermion] (f1),
				(i2) -- [fermion] (b) -- [fermion] (c) -- [fermion] (f2),
				(a) -- [scalar] (b),
				(a) -- [ghost] (c),
				};
				
			\end{feynman}
		
		\end{tikzpicture}
		\label{subfig:EIH_h}
	} \hspace{1em}
	\subfloat[]{
		\begin{tikzpicture}
			\begin{feynman}
				\vertex (i1);
				\vertex [right=of i1] (a);
				\vertex [right=of a] (f1);
				\vertex [below=2em of a] (b);
				\vertex [below=of i1] (i2);
				\vertex [right=2em of i2] (c);
				\vertex [right=of c] (d);
				\vertex [below=of f1] (f2);
				
				\diagram*{
				(i1) -- [fermion] (a) -- [fermion] (f1),
				(i2) -- [fermion] (c) -- [fermion] (d) -- [fermion] (f2),
				(a) -- [scalar] (b) -- [scalar] {(c), (d)},
				};
				
			\end{feynman}
		
		\end{tikzpicture}
		\label{subfig:EIH_i}
	} \hspace{1em}
	\subfloat[]{
		\begin{tikzpicture}
			\begin{feynman}
				\vertex (i1);
				\vertex [right=of i1] (a);
				\vertex [right=of a] (f1);
				\vertex [below=2em of a] (b);
				\vertex [below=of i1] (i2);
				\vertex [right=2em of i2] (c);
				\vertex [right=of c] (d);
				\vertex [below=of f1] (f2);
				
				\diagram*{
				(i1) -- [fermion] (a) -- [fermion] (f1),
				(i2) -- [fermion] (c) -- [fermion] (d) -- [fermion] (f2),
				(a) -- [scalar] (b) -- [ghost] {(c), (d)},
				};
				
			\end{feynman}
		
		\end{tikzpicture}
		\label{subfig:EIH_j}
	} \hspace{1em}
	\subfloat[]{
		\begin{tikzpicture}
			\begin{feynman}
				\vertex (i1);
				\vertex [right=of i1] (a);
				\vertex [right=of a] (f1);
				\vertex [below=2em of a] (b);
				\vertex [below=of i1] (i2);
				\vertex [right=2em of i2] (c);
				\vertex [right=of c] (d);
				\vertex [below=of f1] (f2);
				
				\diagram*{
				(i1) -- [fermion] (a) -- [fermion] (f1),
				(i2) -- [fermion] (c) -- [fermion] (d) -- [fermion] (f2),
				(a) -- [ghost] (b) -- [ghost] (c),
				(b) -- [scalar] (d),
				};
				
			\end{feynman}
		
		\end{tikzpicture}
		\label{subfig:EIH_k}
	} \hspace{1em}
	
\caption{Feynman diagrams contributing to the $Lv^2$ potential. Each diagram that is not symmetric under the exchange of particles wordlines should be added with its symmetric counterpart.}
\label{fig:EIH}
\end{figure}
To order $v^2$, the power counting rules dictate that ten diagrams, shown in Fig.~\ref{fig:EIH},  contribute to the potential. Let us see how each of them contributes. (For notational convenience, since we focus on the Lagrangian, we  remove the $i \int dt$ factor in front of each term.)
\begin{itemize}
\item Figures \ref{subfig:EIH_A} and \ref{subfig:EIH_b} respectively come from the exchange of a potential graviton and scalar with lowest-order vertex and modified propagators (see eqs.~\eqref{eq:modH} and \eqref{eq:modPhi}) and yield
\begin{equation}
\frac{\tilde G_{12} m_1 m_2}{2 r} \left( \mathbf{v}_1 \cdot \mathbf{v}_2 - \frac{(\mathbf{v}_1 \cdot \mathbf{r}) (\mathbf{v}_2 \cdot \mathbf{r})}{r^2} \right) \;.
\end{equation}

\item Figure \ref{subfig:EIH_c} comes from the exchange of a potential graviton with the vertex $h_{0i}v^i$ in the action \eqref{eq:pp_Action_expanded} and yield
\begin{equation}
-4 \frac{G_N m_1 m_2}{r} \mathbf{v}_1 \cdot \mathbf{v}_2 \;.
\end{equation}

\item Figures \ref{subfig:EIH_d} (\ref{subfig:EIH_e})  comes  from the exchange of a potential graviton (scalar) with one of the vertices $hv^2$ ($\phi v^2$) in the action \eqref{eq:pp_Action_expanded} and yields
\begin{equation}
\frac{G_N(3-2 \aaa_1 \aaa_2) m_1 m_2}{2 r} (v_1^2 + v_2^2)\;.
\end{equation}

\item  Figures \ref{subfig:EIH_f}, \ref{subfig:EIH_g} and \ref{subfig:EIH_h}, respectively coming from the $h^2$, $\phi^2$ and $h \phi$ vertices in the action \eqref{eq:pp_Action_expanded}, yield
\begin{equation}
\frac{G_N^2(1 +4 f_{12} -4\aaa_1 \aaa_2) m_1 m_2 (m_1+m_2)}{2 r^2} \;,
\end{equation}
where 
\be
\label{eq:f12}
f_{AB} \equiv \bbb_A \aaa_B^2 + \bbb_B \aaa_A^2 + \kappa_{AB} \left(\bbb_B\aaa_A^2 - \bbb_A \aaa_B^2 \right) \;
\ee
is a symmetric (in the indices $AB$) function built out of $\aaa_A$, $\bbb_A$ and the antisymmetric mass ratio 
\be
\label{eq:kappa}
\kappa_{AB} \equiv \frac{m_A-m_B}{m_A+m_B} \;.
\ee

\item Figure \ref{subfig:EIH_i} comes from the $H^3$ term in the Einstein-Hilbert action and yields
\begin{equation}
-\frac{G_N^2 m_1 m_2 (m_1+m_2)}{r^2} \;.
\end{equation}

\item Finally, the last diagrams in Figs.~\ref{subfig:EIH_j} and \ref{subfig:EIH_k} do not contribute because they are proportional to $\frac{\eta^{\alpha \beta}}{2} P_{00,\alpha \beta} \delta_{ij} - P_{00,ij}$, which vanishes.
\end{itemize}

Gathering all these terms, we can put the action into the following form 
\be
\begin{split}
\label{eq:LEIH}
S_{Lv^2} =& \int dt \;  \bigg\{ \frac{1}{8} \sum_A m_A v_A^4 \\
&+  \frac{\tilde{G}_{12} m_1 m_2}{2 r} \left[ (v_1^2 + v_2^2) - 3 \mathbf{v}_1 \cdot \mathbf{v}_2 -  \frac{(\mathbf{v}_1 \cdot \mathbf{r}) (\mathbf{v}_2 \cdot \mathbf{r})}{r^2} + 2\gamma_{12} (\mathbf{v}_1 - \mathbf{v}_2)^2 \right] \\
& - \frac{\tilde{G}_{12}^2 m_1 m_2 (m_1+m_2)}{2 r^2} (2 \beta_{12} -1) \bigg\} \;,
\end{split}
\ee
where as usual  $\tilde{G}_{12} = G_N (1+2 \aaa_1 \aaa_2)$ is the effective Newton constant and $\beta_{AB}$ and $\gamma_{AB}$ are PPN parameters
given here by\footnote{Analgous parameters are defined in the context of multi-scalar-tensor gravity in Ref.~\cite{damour_tensor-multi-scalar_1992}, see eq.~(6.26). The dictionary between our and their notation  is $\alpha^{\rm there}_A = - \aaa^{\rm here}_A \sqrt{2}$, $\beta^{\rm there}_{A} = -4 \bbb^{\rm here}_A -2 (\aaa^{\rm here}_A)^2$.}
\begin{align}
\gamma_{AB} &= 1 - 4 \frac{\aaa_A \aaa_B}{1+2\aaa_A \aaa_B} \;, \\
\beta_{AB} &= 1 - 2 \frac{\aaa_A^2\aaa_B^2 + f_{AB}}{(1+2\aaa_A \aaa_B)^2} \;.
\end{align}
For $\aaa_A=0=\bbb_A$ we recover the EIH correction to the Newtonian dynamics originally derived in \cite{einstein_gravitational_1938} 
and reproduced in the the framework of the NRGR approach in \cite{goldberger_effective_2006,Kol:2007bc}. 
More generally, the above action agrees with that of Ref.~\cite{damour_tensor-multi-scalar_1992} (see eq.~(3.7) of that reference).\footnote{A similar calculation has been done in \cite{Huang:2018pbu}  for a massive axion-type field. However, their result disagrees with ours (and with \cite{damour_tensor-multi-scalar_1992}) in the massless limit. The disagreement may be traced in the calculation of the diagrams in Figs.~\ref{subfig:EIH_j} and \ref{subfig:EIH_k}. We thank the authors of Ref.~\cite{Huang:2018pbu} for double checking their results and eventually agreeing with us in a private correspondence.}

\section{Couplings to radiative fields}
 \label{sec:couplings} 

In this section we compute the couplings of the radiated fields to the point particles up to 2.5PN order.
In general, we can expand the effective action $S_\mathrm{eff}[x_A, \bar{h}_{\mu \nu}, \bar{\varphi}]$ as
\begin{equation}
S_\mathrm{eff}[x_A,  \bar{h}_{\mu \nu}, \bar{\varphi}] = S_0[x_A] + S_1[x_A,  \bar{h}_{\mu \nu}, \bar{\varphi}] + S_2 [x_A,  \bar{h}_{\mu \nu}, \bar{\varphi}]  + S_{\rm NL} [x_A,  \bar{h}_{\mu \nu}, \bar{\varphi}]  \;.
\end{equation}
The first term of the right-hand side, $S_0$, does not depend on external radiation gravitons. This is the conservative part of the action  that we have computed in Section \ref{sec:EIH_Lagrangian}  and can be discarded from the following discussion. The next term, $S_1$, is linear in the radiating fields and contains the source that the radiating fields are coupled to. 
On  general grounds, it can be written as 
\begin{equation}
\label{eq:S1}
S_1 = S_{\rm int}^{(h)} + S_{\rm int}^{(\varphi)} \;, \quad S_{\rm int}^{(h)} \equiv -\frac{1}{2\Mp} \int d^4x T^{\mu \nu}(x) \bar{h}_{\mu \nu}(x) \;, \quad S_{\rm int}^{(\varphi)} \equiv \frac{1}{\Mp} \int d^4x J(x) \bar{\varphi}(x)\;, 
\end{equation}
where $T^{\mu \nu}$ and $J$ are respectively the sources for the metric and  the  scalar field radiation fields. In particular, $T^{\mu \nu}$ is the (pseudo) matter  energy-momentum tensor that includes the gravitational self-energy---i.e.~the contributions from the integrated out potential gravitons. It  is conserved in flat spacetime, $\partial_\mu T^{\mu \nu} = 0$, by linear diffeomorphism invariance.

The part quadratic in the radiating fields, $S_2$, provides the kinetic terms of $\bar{h}_{\mu \nu}$ and $\bar \varphi$ while $S_{\rm NL}$ contains higher-order coupling terms. 
The non-linear couplings  in the radiating fields give rise to the so-called \textit{tail effects} \cite{porto_effective_2016, goldberger_gravitational_2010} and will not be discussed here because they are of order   $1.5$PN higher than the leading order  quadrupole.

Following \cite{ross_multipole_2012}, to discuss the  couplings to the radiation fields and highlight the power counting in $v$ of the emission process~\cite{goldberger_effective_2006},
we will perform a multipole expansion of the sources of  $\bar{h}_{\mu \nu}$  and $\bar{\varphi}$   at the level of the action.
To simplify the treatment, we will focus here only on the lowest-order coupling but the full derivation can be found in  \cite{ross_multipole_2012, goldberger_gravitational_2010}.
We will quickly review the graviton case, which has been discussed at length in the literature \cite{goldberger_gravitational_2010}. We will turn in more details to the scalar case below.

\subsection{Graviton interactions}

\subsubsection{Multipole decomposition} 
Let us first consider the coupling of radiation gravitons with the sources, $S_{\rm int}^{(h)}$. In this subsection we consider a general stress-energy tensor $T^{\mu \nu}$.
In the next subsection we will give the explicit expression of $T^{\mu \nu}$ for our particular physical configuration.

To simplify the calculation and because this case has been studied at length in many references (see e.g.~\cite{goldberger_effective_2006,galley_radiation_2009}\footnote{If one does not chose this gauge and keeps all the components in the discussion, one finds that $\bar{h}_{00}$ couples to the total mass and Newtonian energy of the system while $\bar{h}_{0i}$ couples to the leading-order orbital angular momentum. These are conserved quantities at Newtonian order, which implies that they do not contribute to the radiation emission. 
Moreover, one can find that the quadrupole moment of the stress-energy tensor $I_h^{ij} $ couples to the linearized ``electric-type'' part of the Riemann tensor, $R_{0i0j}$, given by \be
R_{0i0j} \equiv - \frac{1}{2\Mp} \big(\partial_i \dot {\bar h}_{0j} + \partial_j \dot {\bar h}_{0i} - \ddot { \bar{h}}_{ij} - \partial_i \partial_j \bar h_{00} \big)\;,
\ee
whose two-point function is proportional to the projection operator into symmetric and traceless two-index spatial tensors.}), we directly focus on the so-called transverse-traceless gauge, defined by 
\be
\bar h_{0\mu} = 0  \;, \qquad \partial^i \bar h_{ji} =0 \;, \qquad  \bar h^k_k =0 \;.
\ee
Denoting by $\bar{h}_{ij}^{\rm TT}$  the radiated graviton in this gauge, the graviton interaction vertex of eq.~\eqref{eq:S1}  is 
\begin{equation}
\label{eq:gravcoup}
S_{\rm int}^{(h)} = -\frac{1}{2\Mp} \int d^4x T^{ij} \bar{h}_{ij}^{\rm TT} \;.
\end{equation}
Using the equation of motion $\partial_\mu T^{\mu \nu} = 0$, it is straightforward  to rewrite this equation as 
\be
\label{eq:interaction_graviton_multipole}
S_{\rm int}^{(h)} = - \frac{1}{2} \int dt I_h^{ij}  \frac{1}{2\Mp} \ddot { \bar{h}}_{ij}^{\rm TT} \;,  
\ee
where $I_h^{ij}$ is the quadrupole moment of the stress-energy tensor, defined as
\be
\label{eq:Ih}
I_h^{ij} \equiv \int d^3x T^{00} \left( x^i x^j - \frac{1}{3}x^2 \delta^{ij}\right) \;.
\ee

\subsubsection{Quadrupole expression}

As we have just seen in eq.~\eqref{eq:Ih},  to find the gravitational interaction vertex up to quadrupole order we just needed $T^{00}$ to lowest order. After comparison with the full action, eq.~\eqref{eq:pp_Action_expanded}, this is given by 
\begin{equation}
T^{00} = - \sum_A m_A \delta^3(\mathbf{x}-\mathbf{x}_A) \;,
\end{equation}
and the expression of the lowest-order quadrupole is the usual one, i.e.,
\begin{equation}
\label{eq:quadrupole_gr}
I_h^{ij} = - \sum_A m_A  \left( x_A^i x_A^j - \frac{1}{3}x_A^2 \delta^{ij}\right) \;.
\end{equation}
Therefore, at this order the vertex \eqref{eq:gravcoup} is not modified by the presence of the scalar. However, as we will discuss below, to compute the emitted power we will have to take the third derivative of  the quadrupole moment with respect to time, see eq.~\eqref{eq:quadrupole_graviton}. This involves the acceleration of the two bodies and thus, using the equations of motion,  the modified Newton constant $\tilde{G}_{12} = G_N (1+2 \aaa_1 \aaa_2)$.
Note that, by using the NRGR power-counting rules explained in Sec. \ref{sec3}, one finds that the gravitational quadrupole interaction vertex of eq. \eqref{eq:interaction_graviton_multipole} is of order $\sqrt{L v^5}$.

\subsection{Scalar interactions}

\subsubsection{Multipole decomposition}
\label{subsubsec:multipoledec}

Let us now consider the coupling of radiation scalars with the sources, $S_{\rm int}^{(\varphi)}$.
Including also the  quadratic action of the radiating scalar,
we have
\be
S_{\rm eff} \supset S_2^{(\varphi)} + S_{\rm int}^{(\varphi)}=  \int d^4x  \left( -\frac{1}{2} \eta^{\mu \nu}  \partial_\mu \bar{\varphi} \partial_\nu \bar{\varphi} 
+ \frac{1}{\Mp} J   \bar{\varphi} \right)  \;, 
\ee
which leads to the following equation of motion 
\be
\square \bar{\varphi} = -\frac{J}{\Mp} \;, \qquad \square \equiv \eta^{\mu \nu}  \partial_\mu  \partial_\nu \;.
\ee

Now we want to do a multipole expansion of the scalar field around the center-of-mass $\mathbf{x}_{\rm cm}$, which is defined by\footnote{This definition comes from the invariance of the theory under boosts, which  via Noether theorem gives that the following charge is conserved, 
\begin{equation}
Q^{0i} = \int d^3x \left( T^{00} x^i - T^{0i} t \right) \;.
\end{equation}
Since the total momentum $P^i = \int d^3x T^{0i}$ and energy $E = \int d^3x T^{00}$ are also conserved, we get that the center-of-mass moves with a constant velocity, thus justifying its definition.
Even for standard gravity and point-particle masses, since there are higher-order corrections implied by eq.~\eqref{eq:center_of_mass} the definition $ x_{\rm cm}^i \equiv \sum_A m_A x_A^i/\left( \sum_A m_A \right)$ is valid only at lowest order in the velocity expansion.} 
\begin{equation} \label{eq:def_center_mass}
 \mathbf{x}_{\rm cm} \equiv \frac1{E} \int d^3x \, T^{00} \mathbf{x} \;, \qquad E \equiv  \int d^3 x \, T^{00} \;,
\end{equation}
and can be set at the origin without loss of generality, $\mathbf{x}_{\rm cm}= \mathbf{0}$.
Since we are considering the physical configuration where the radiating scalar field $\bar \varphi$ varies on scales that are much larger than the source term $J$, we can expand the scalar in the interaction $S_{\rm int}^{(\varphi)}$ defined in eq.~\eqref{eq:S1} around the center of mass.  This gives
\begin{equation}
\begin{split}
S_{\rm int}^{(\varphi)} = \int d^4 x \frac{J \bar \varphi }{\Mp}= \int dt \, \int d^3 x \frac{J (t, \mathbf{x})}{\Mp}  \bigg( \bar{\varphi}(t, \mathbf{0}) + x^i \partial_i \bar{\varphi}(t,\mathbf{0}) + \frac{1}{2} x^i x^j \partial_i \partial_j \bar{\varphi}(t, \mathbf{0})  \\
+ \frac{1}{3!} x^i x^j x^k \partial_i \partial_j \partial_k \bar{\varphi}(t, \mathbf{0}) + \ldots \bigg) \;,
\end{split}
\end{equation}
which allows to obtain an expansion of the interactions in terms of the moments of the source, $\int d^3 x {J (t, \mathbf{x}) x^n}$, $n=0,1,\ldots$. 
Recalling that $\partial_i \bar{\varphi} \sim ({v}/{r}) \bar \varphi$, each moment $n$ enters suppressed by $v^n$.

This is not yet organised as a multipole expansion. To achieve this, instead of the moments one should use their irreducible representations under the rotation group. The first term to be modified is the second moment. Instead of $x^i x^j$ we should use 
\be
Q^{ij} \equiv x^i x^j - \frac{1}{3}x^2 \delta^{ij}\;. 
\ee
To compensate the additional term added, one is left with  
\begin{equation}
\frac{1}{6\Mp} \int d^3x \, x^2 J \, \nabla^2 \bar{\varphi} \;.
\end{equation}
We can then use the scalar equation of motion that, up to a contact term renormalizing the point-particle masses, transforms this term into a monopole one. 
Similarly, the  symmetric and traceless tensor associated to the third moment is 
\begin{equation}
Q^{ijk} = x^i x^j x^k - \frac{1}{5} \left(\delta^{ij} x^2 x^k + 2 \; \mathrm{perm} \right) \; ,
\end{equation}
which generates the term 
\begin{equation}
\frac{1}{10\Mp} \int d^3x \, x^2 x^i J \, \partial_i \nabla^2 \bar{\varphi} \;,
\end{equation}
which upon use of the scalar equation of motion contributes to the dipole.
This can go on but we only need these terms for the order we are considering.

Finally, this gives for the interaction (up to order $v^2$),
\begin{align}
\begin{split}
S_{\rm int}^{(\varphi)} =  \frac{1}{\Mp} \int dt \left( I_\varphi \bar{\varphi} + I_\varphi^i \partial_i \bar{\varphi} + \frac{1}{2} I_\varphi^{ij} \partial_i \partial_j \bar{\varphi}  + \ldots \right) \;,
\end{split}
\label{eq:multipole_expansion_scalar}
\end{align}
where 
\be
\label{eq:monodipoquadru}
I_\varphi  \equiv  \int d^3 x \left( J + \frac{1}{6}  \partial_t^2 J  x^2 \right) \;, \quad I_\varphi^i  \equiv \int d^3x \, x^i \left(J + \frac{1}{10} \partial_t^2 J x^2 \right) \;, \quad  I_\varphi^{ij} \equiv \int d^3x J  Q^{ij} \;
\ee
are respectively the scalar monopole, dipole and quadrupole.

\subsubsection{Scalar monopole}

Let us discuss the coupling induced by the monopole, i.e.~the first term on the right-hand side of eq.~\eqref{eq:multipole_expansion_scalar}. At lowest order in $v$, the source is 
\begin{equation}
J_{v^0} = \sum_A \aaa_A m_A \delta^3(\mathbf{x}-\mathbf{x}_A) \;,
\label{eq:J0}
\end{equation}
which translates into a scaling 
\begin{equation}
S_\mathrm{mono}^{(\varphi)} \sim \frac{I_\varphi}{\Mp} \sim \sum_A \aaa_A \sqrt{Lv} \;.
\end{equation}
This could be potentially very constraining if compared to the  radiation graviton, which starts at the quadrupole order of $\sqrt{L v^5}$.
However, this gives a constant coupling and thus,  as we will see in Sec.~\ref{subsec:radiated_scalars}, eq.~\eqref{eq:radiated_power_scalar}, no scalar radiation is emitted.  Therefore, we need to go to higher order.

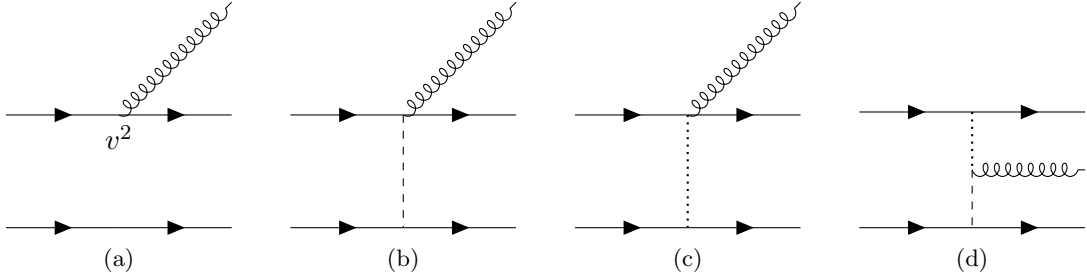
\begin{figure}
	\centering
	\subfloat[]{
		\begin{tikzpicture}
			\begin{feynman}
				\vertex (i1);
				\vertex [right=of i1] (a);
				\vertex [right=of a] (f1);
				\vertex [above=of f1] (s);
				\vertex [below=of i1] (i2);
				\vertex [below=of a] (b);
				\vertex [below=of f1] (f2);
				
				\diagram*{
				i1 -- [fermion] (a) -- [fermion] (f1),
				(a) -- [gluon] (s),
				(i2) -- [fermion] (b) -- [fermion] (f2),
				};
				
				\draw[] (a.south) node[below] {$v^2$};
			\end{feynman}
		
		\end{tikzpicture}
		\label{subfig:J_v2_A}
	}\hspace{1em}
	\subfloat[]{
		\begin{tikzpicture}
			\begin{feynman}
				\vertex (i1);
				\vertex [right=of i1] (a);
				\vertex [right=of a] (f1);
				\vertex [above=of f1] (s);
				\vertex [below=of i1] (i2);
				\vertex [below=of a] (b);
				\vertex [below=of f1] (f2);
				
				\diagram*{
				i1 -- [fermion] (a) -- [fermion] (f1),
				(a) -- [gluon] (s),
				(i2) -- [fermion] (b) -- [fermion] (f2),
				(a) -- [scalar] (b)
				};
			\end{feynman}
		
		\end{tikzpicture}
		\label{subfig:J_v2_b}
	}\hspace{1em}
	\subfloat[]{
		\begin{tikzpicture}
			\begin{feynman}
				\vertex (i1);
				\vertex [right=of i1] (a);
				\vertex [right=of a] (f1);
				\vertex [above=of f1] (s);
				\vertex [below=of i1] (i2);
				\vertex [below=of a] (b);
				\vertex [below=of f1] (f2);
				
				\diagram*{
				i1 -- [fermion] (a) -- [fermion] (f1),
				(a) -- [gluon] (s),
				(i2) -- [fermion] (b) -- [fermion] (f2),
				(a) -- [ghost] (b)
				};
			\end{feynman}
		
		\end{tikzpicture}
		\label{subfig:J_v2_c}
	}\hspace{1em}
	\subfloat[]{
		\begin{tikzpicture}
			\begin{feynman}
				\vertex (i1);
				\vertex [right=of i1] (a);
				\vertex [right=of a] (f1);
				\vertex [below=2em of f1] (s);
				\vertex [below=2em of a] (i);
				\vertex [below=4em of i1] (i2);
				\vertex [below=2em of i] (b);
				\vertex [below=4em of f1] (f2);
				
				\diagram*{
				i1 -- [fermion] (a) -- [fermion] (f1),
				(i) -- [gluon] (s),
				(i2) -- [fermion] (b) -- [fermion] (f2),
				(a) -- [ghost] (i) -- [scalar] (b)
				};
			\end{feynman}
		
		\end{tikzpicture}
		\label{subfig:J_v2_d}
	}\hspace{1em}

\caption{Feynman diagrams contributing to the emission of one scalar, at  order $v^2$. Diagrams that are not symmetric should be added with their symmetric counterpart.}
\label{fig:J_v2}
\end{figure}

Integrating out potential gravitons and scalars, we find  four diagrams that contribute to $J$ at order $\sqrt{Lv^5}$, all shown in Fig.~\ref{fig:J_v2}. 
The expression for these diagrams are (for convenience we suppress the $i \int dt$ in front of each diagram):
\begin{itemize}

\item Figure \ref{subfig:J_v2_A}, coming from the $v^2$ term in $\int d\tau \varphi$ (see eq.~\eqref{eq:pp_Action_expanded}),
\begin{equation}
- \sum_A \aaa_A \frac{m_A v_A^2}{2} \frac{\bar{\varphi}(t, \mathbf{x}_A)}{\Mp} \;.
\end{equation}

\item Figure \ref{subfig:J_v2_b}, coming from the $\bar \varphi H$ term in $\int d\tau \varphi$,\footnote{In \cite{Huang:2018pbu}, a similar calculation of this diagram (denoted by $7b$ there), for a massive axion-type field, has been reported. We disagree with their result in the massless limit.}
\begin{equation}
- \frac{m_1 m_2 G_N}{r} \sum_A \aaa_A \frac{\bar{\varphi}(t, \mathbf{x}_A)}{\Mp}\;.
\end{equation}

\item Figure \ref{subfig:J_v2_c}, coming from the $\bar{\varphi} \Phi$ term in $\int d\tau \varphi^2$,
\begin{equation}
4 \frac{m_1 m_2 G_N}{r}  \sum_A \bbb_A \aaa_{\bar{A}} \frac{\bar{\varphi}(t, \mathbf{x}_A)}{\Mp}\;,
\end{equation}
where for compactness we have introduced the notation $\aaa_{\bar{A}}$ for the symmetric parameter, i.e.~$\aaa_{\bar{1}} = \aaa_2$ and $\aaa_{\bar{2}}=\aaa_1$.

\item Figure \ref{subfig:J_v2_d}, coming from the ${\varphi} \Phi H$ term of eq.~\eqref{eq:hPhi2Vertex}. This  vanishes as in the conservative case, because it involves the same projector $\frac{\eta^{\alpha \beta}}{2} P_{00,\alpha \beta} \delta_{ij} - P_{00,ij}$.

\end{itemize}
In conclusion, the complete expression for the coupling $J$ at  order $v^2$ is 
\begin{equation}
J_{v^2} = - \sum_A m_A \aaa_A \frac{v_A^2}{2} \delta^3(\mathbf{x}-\mathbf{x}_A) + \frac{m_1 m_2 G_N}{r} \sum_A (4 \bbb_A \aaa_{\bar{A}} - \aaa_A) \delta^3(\mathbf{x}-\mathbf{x}_A) \;.
\label{eq:J_v2}
\end{equation}

Let us now discuss the second term in $I_\varphi $, i.e.~$\int d^3x \frac{1}{6} (\partial_t^2 J_{v^0}) x^2$. To calculate it, we use the equations of motion for the point-particles  at lowest order in the velocity expansion, i.e.,
\be
\mathbf{\ddot{x}}_1 = - \frac{\tilde G_{12} m_2}{r^3} \mathbf{r} \;,  \qquad
\mathbf{\ddot{x}}_2 =  \frac{\tilde G_{12} m_1}{r^3} \mathbf{r} \;,
\ee
with the following center-of-mass relations, also valid  at lowest order in the velocity expansion,
\be
\mathbf{x}_1 = \frac{m_2}{m_1+m_2} \mathbf{r} \;, \qquad
\mathbf{x}_2 = - \frac{m_1}{m_1+m_2} \mathbf{r}  \;.
\label{eq:center_of_mass}
\ee

Summing up all contributions and using eq.~\eqref{eq:monodipoquadru}, we finally find
\begin{equation}
I_\varphi = - \frac{1}{6} \sum_A m_A \aaa_A v_A^2 + g_{12} \frac{G_N m_1 m_2}{r} \;,
\end{equation}
where  $g_{AB}$ is a symmetric combination of the scalar couplings $\aaa_A$, $\bbb_A$ and of the antisymmetric mass ratio $\kappa_{AB}$ defined in eq.~\eqref{eq:kappa}, \begin{equation}
g_{AB} \equiv \aaa_A(4\bbb_B-1)+\aaa_B(4\bbb_A-1) - \frac{1+2\aaa_A\aaa_B}{6}(\aaa_A+\aaa_B+\kappa_{AB}(\aaa_B-\aaa_A)) \;.
\end{equation}

\subsubsection{Scalar dipole}

The fact that scalar-tensor theories generically predict a dipole was first realized by Eardley \cite{Eardley1975ApJ}. This could induce sizeable deviations from GR because the scalar dipole interaction term is \textit{a priori} of order $\sqrt{Lv^3}$. Using in eq.~\eqref{eq:monodipoquadru} the lowest-order expression for $J$ (eq.~\eqref{eq:J0}), we obtain the lowest-order contribution to the dipole, 
\begin{equation}
I_{\varphi, \rm -1PN}^i = \sum_A \aaa_A m_A x_A^i  \;.
\label{eq:scalar_dipole}
\end{equation}
For equal scalar charges of the two objects,  $\aaa_1 = \aaa_2$,  
the second derivative of $I^i_\varphi$ vanishes due to the conservation of the total momentum and there is no $-$1PN dipole radiation for equal scalar charges.
Thus, for two black holes or two comparable neutron stars, the effect of the dipole is very weak, while for a black hole-neutron star system it is maximal (a black hole has $\aaa_{\rm BH}=0$ in traditional scalar-tensor theories due to the no-hair theorem). See~\cite{Huang:2018pbu} for a detailed discussion.

Let us now compute the first-order (1PN) correction to this expression. To do that, we have to take into account the second term in $I^i_\varphi $  in eq.~\eqref{eq:monodipoquadru} coming from the trace part of the octupole, i.e.~$\int d^3x x^i \frac{1}{10} (\partial_t^2 J_{v^0}) x^2$, and the $v^2$ correction to the source from eq.~\eqref{eq:J_v2}. Adding these to the leading-order expression above gives
\begin{align} \label{eq:dipole_full}
\begin{split}
I_\varphi^i &= \sum_A \aaa_A m_A x_A^i + \frac{1}{10} \partial_t^2 \left(\sum_A m_A \aaa_A x_A^2 x_A^i \right) - \sum_A m_A \aaa_A \frac{v_A^2}{2} x_A^i \\
&+ \frac{G_N m_1 m_2}{r} \sum_A (4 \beta_A \aaa_{\bar{A}} - \aaa_A) x_A^i \;.
\end{split}
\end{align}

\subsubsection{Scalar quadrupole}

The scalar quadrupole interaction vertex is of order $\sqrt{Lv^5}$ and can be straightforwardly computed from eq.~\eqref{eq:monodipoquadru} and the lowest-order expression for $J$. One finds
\begin{equation}
I_\varphi^{ij} =  \sum_A \aaa_A m_A \left(  x_A^i x_A^j - \frac{1}{3}x_A^2 \delta^{ij} \right) \;.
\label{eq:scalar_quadrupole}
\end{equation}

\section{Dissipative dynamics \label{sec:dissipative_dynamics}}

Now that we have a definite expansion for the interaction Lagrangian in terms of multipole moments, we can calculate the power emitted in gravitational waves. As explained below, this can be computed from the imaginary part of the effective action for the two point-like bodies, $\hat S_\mathrm{eff}[x_A]$, obtained by integrating out the radiation fields, see eq.~\eqref{eq:finalSeff}. 

The real part of the effective action generates the coupled equations of motion for the two-body system.
If some energy leaves the system, then $\hat  S_\mathrm{eff}$ contains an imaginary part that is related to the power emitted. To see why this is the case by a simple example, we consider 
a scalar theory with a field $\phi$ coupled to an external source $J$ entering the action as  $\int d^4x J(x) \phi(x)$. 
The effective action obtained by integrating the field $\phi$ is given by the path integral
\begin{equation}\label{pathint}
e^{i S_{\rm eff}[J]}  = \int \mathcal{D}\phi \, e^{iS[\phi, J]} \equiv  Z[J] \;,
\end{equation}
where in the last equality we have defined the generating functional of the Green's functions $Z[J]$.
On the other hand, in the so-called ``in-out'' formalism, $Z$ is also the overlap between initial and final states, i.e.
\begin{equation}
Z[J] = \braket{0_+}{0_-}_J \;.
\end{equation}
From the two equations above, the vacuum transition amplitude between the asymptotic past and future differs from unity if the effective action is not real,
\begin{equation}
|\braket{0_+}{0_-}_J|^2 = e^{-2 \Im[S_{\rm eff}]}  \;.
\label{eq:trans}
\end{equation}
The difference with unity denotes the probability amplitude that particles are lost---or emitted---by the system. 
Expanding the right-hand side of this equation for small $\Im[S_{\rm eff}]$, this can be written as  
\begin{equation}
2 \Im[S_{\rm eff}] = T \int dE d\Omega \frac{d^2 \Gamma}{dEd\Omega} \;,
\label{eq:imaginaryPartAction}
\end{equation}
where $T$ is the  duration of the interaction and $d \Gamma$ is the differential rate for particle emission. The latter can be employed to calculate the radiated power via 
\be
P = \int dE d\Omega E \frac{d^2 \Gamma}{dEd\Omega} \;.
\label{eq:power}
\ee
We will use the two equations above to compute the power radiated into gravitons and scalar particles. 

\subsection{Radiated power}
\subsubsection{Gravitons}
\label{subsec:radiated_gravitons}

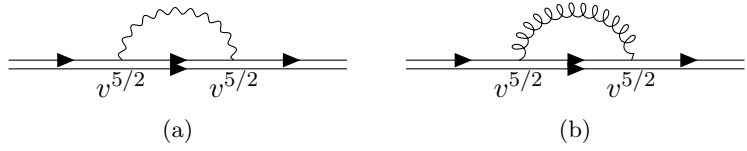
\begin{figure} 
	\centering
	\subfloat[]{
		\begin{tikzpicture}
			\begin{feynman}
				\vertex (i1);
				\vertex [right=of i1] (a);
				\vertex [right=of a] (b);
				\vertex [right=of b] (f1);
				\vertex [below=0.3em of i1] (i2);
				\vertex [below=0.3em of f1] (f2);
				
				\diagram*{
				i1 -- [fermion] (a) -- [fermion] (b) -- [fermion] (f1),
				(i2) -- [fermion] (f2),
				(a) -- [photon, half left] (b)
				};
				
				\draw[] (a.south) node[below] {$v^{5/2}$};
				\draw[] (b.south) node[below] {$v^{5/2}$};
			\end{feynman}
		
		\end{tikzpicture}
		\label{subfig:radiation_grav_Seff}
	} \hspace{1em}
	\subfloat[]{
		\begin{tikzpicture}
			\begin{feynman}
				\vertex (i1);
				\vertex [right=of i1] (a);
				\vertex [right=of a] (b);
				\vertex [right=of b] (f1);
				\vertex [below=0.3em of i1] (i2);
				\vertex [below=0.3em of f1] (f2);
				
				\diagram*{
				i1 -- [fermion] (a) -- [fermion] (b) -- [fermion] (f1),
				(i2) -- [fermion] (f2),
				(a) -- [gluon, half left] (b)
				};
				
				\draw[] (a.south) node[below] {$v^{5/2}$};
				\draw[] (b.south) node[below] {$v^{5/2}$};
			\end{feynman}
		
		\end{tikzpicture}
		\label{subfig:radiation_scalar_Seff}
	} \hspace{1em}
	
\caption{Contribution of the radiation graviton and scalar to the imaginary part of the effective action.}
\label{fig:radiation_Seff}
\end{figure}
Let us first compute the power radiated into gravitons (see e.g.~\cite{goldberger_effective_2006,Kol:2007bc,galley_radiation_2009,Cardoso:2002pa}).  
In the classical approximation the path-integral~\eqref{pathint} is computed at the saddle point of the action,  $\hat S_{\rm eff}[x_A] = S_{\rm eff}[x_A, h_{cl}, \varphi_{cl}]$, and thus decomposes into the two diagrams of~Fig.~\ref{fig:radiation_Seff}: $\hat S_{\rm eff}=\hat S_{\rm eff}^{(h)}+ \hat S_{\rm eff}^{(\varphi)}$. The first term (Fig.~\ref{subfig:radiation_grav_Seff}), contains the interaction vertex of eq.~\eqref{eq:interaction_graviton_multipole}. In particular, using the Feynman rules from this equation we find 
\be
i  \hat S_{\rm eff}^{(h)} = - \frac12 \times \frac{1}{16 \Mp^2} \int dt_1 dt_2   I_h^{ij}(t_1) I_h^{kl}(t_2) \left\langle T \ddot {\bar h}_{ij}^{\rm TT} (t_1, \mathbf{0})   \ddot {\bar h}_{kl}^{\rm TT} (t_2, \mathbf{0}) \right\rangle \;,
\label{eq:Seffgrav}
 \ee
 where we have included the symmetry factor $1/2$ of the diagram.

 To find the propagator for $\bar{h}_{ij}^{\rm TT}$, we can first project $\bar{h}_{ij}$ on the transverse-traceless gauge.
 In terms of the unit vector $\mathbf{ n}$ denoting the direction of  propagation, we have
 \be
 \bar{h}_{ij}^{\rm TT} =  \Lambda_{ij,kl}  ({\mathbf{n}} ) \bar{h}_{kl} \;, 
 \ee 
 where 
 \be
 \Lambda_{ij,kl}  ({\mathbf{n}} ) \equiv  (\delta_{ik}-n_in_k)(\delta_{jl}-n_jn_l)-\frac{1}{2}(\delta_{ij}-n_in_j)(\delta_{kl}-n_kn_l) \;,
 \ee
 is the projector---it satisfies $\Lambda_{ij,kl} \,\Lambda_{kl,mn} = \Lambda_{ij,mn}$---onto transverse-traceless tensors, in the sense that it is transverse to $\mathbf{n}$ in all its indices and traceless in the $ij$, $kl$ indices.

Using the  $\bar{h}_{\mu \nu}$ propagator given by eqs.~\eqref{eq:propagator} and \eqref{eq:Feyprop}, applying  the identities
\be
 \label{eq:ur}
\langle n_i n_j \rangle \ = \ \frac13  \delta_{ij}\;, \qquad \langle n_i n_j n_k n_l \rangle\  = \ \frac1{15}  (\delta_{ij} \delta_{kl} + \delta_{ik} \delta_{jl} + \delta_{il} \delta_{jk} ) \;,
\ee
which follow from the rotational symmetry of the integral, and symmetrizing over the indices $ij$ and $kl$, the expectation value in eq.~\eqref{eq:Seffgrav} can be written as 
 \be
\langle  T \ddot{ \bar{h}}_{ij}^{\rm TT} (t_1) \ddot{ \bar{h}}_{kl}^{\rm TT} (t_2) \rangle = \frac85 \left[ \frac12 (\delta_{ik} \delta_{jl} +\delta_{il} \delta_{jk} ) - \frac13 \delta_{ij} \delta_{kl}  \right]  \int  \frac{d^4 k}{(2\pi)^4}    \frac{-i (k_0)^4 }{k^2 - i \epsilon}  e^{i k_0 (t_1-t_2) } \;,
\ee
where the bracket on the right-hand side contains the projection operator
into symmetric and traceless two-index spatial tensors.
Plugging this expression into eq.~\eqref{eq:Seffgrav}, we find
\be
\hat S_{\rm eff}^{(h)} =  \frac{1}{20 \Mp^2} \int  \frac{d^4 k}{(2\pi)^4}    \frac{(k_0)^4}{k^2 - i \epsilon} | I_h^{ij}(k_0) |^2   \;,
 \ee
 where we have introduced the Fourier transform of the quadrupole moment, 
\be
I_h^{ij}(k_0) = \int dt I_h^{ij} (t) e^{i k_0 t}  \;.
\ee

To extract the imaginary part of the above action, we use the relation for the principal value of a function (denoted by PV). Specifically, we have
\begin{equation}
\frac{1}{k^2-i\epsilon} = \text{PV} \left( \frac{1}{k^2} \right) + i\pi \delta(k^2) \;, 
\label{eq:ImEpsilon}
\end{equation}
where 
\be
\delta(k^2) = \frac{1}{2 |k_0|} \left[ \delta(k_0 - |\mathbf{k}|) + \delta(k_0 + |\mathbf{k}|) \right] \;.
\label{eq:deltarel}
\ee
Using these relations, the imaginary part of the effective action reads
\be
{\Im}[\hat {S}_\mathrm{eff}^{(h)}]  = \frac{1}{20\Mp^2} \int \frac{d^3\mathbf{k}}{(2\pi)^3} \frac{|\mathbf{k}|^4}{2 |\mathbf{k}|}   | I_h^{ij}(|\mathbf{k}|)|^2 =\frac{ G_N}5 \int_0^\infty  \frac{d\omega \, \omega^5}{2 \pi}   | {I_h^{ij}}(\omega)|^2    \;,
 \ee
where for the second equality we have  integrated over the angles, used $1/\Mp^2 = 8 \pi G_N$, the on-shell condition $|\mathbf{k}| = \pm k_0$ and defined the emitted frequency as $\omega \equiv | k_0|$.
Comparing with eq.~\eqref{eq:imaginaryPartAction} and applying eq.~\eqref{eq:power}, the expression for the emitted power into gravitons  is 
\be
\begin{split} 
P_g &= \frac{ 2 G_N}{5  T}  \int_0^\infty \frac{d \omega \, \omega^6}{2 \pi} |{I}_h^{ij} (\omega)|^2 \\
&= \frac{G_N}{5 T }
 \int_{-\infty}^\infty dt \dddot{I}_h^{ij} (t) \dddot{I}_h^{ij} (t) \equiv \frac{G_N}{5 }  \big\langle \dddot{I}_h^{ij}\  \dddot{I}_h^{ij} \big\rangle \;,
\label{eq:quadrupole_graviton}
\end{split}
\ee
where in  the second line we have Fourier transformed back the multipoles  to real space and in the last equality we have used the brackets to denote the time average over many gravitational wave cycles.

\subsubsection{Scalars}
\label{subsec:radiated_scalars}

Let us turn now to the power radiated into scalars.
We will now calculate the imaginary part of the effective action $\hat S_{\rm eff}^{(\varphi)}$ obtained by integrating out the radiation scalars. This can be done by computing the self-energy diagram of Fig.~\ref{subfig:radiation_scalar_Seff}, the interaction vertices being the ones of eq.~\eqref{eq:multipole_expansion_scalar}.  
Note that the two vertices in Fig.~\ref{subfig:radiation_scalar_Seff} must be of the same multipole order---if they are not, the remaining indices should be contracted with rotationally invariant tensors, e.g.~$\delta_{ij}$ or $\epsilon_{ijk}$, but such expressions vanish because of the symmetry and the tracelessness of the multipole moments. By applying the multipole expansion derived in Sec.~\ref{subsubsec:multipoledec} and using the Feynman  rules, we get
\be
\begin{split}
i  \hat S_\mathrm{eff}^{(\varphi)} = & - \frac12 \times \frac{1}{\Mp^2} \int dt_1 dt_2 \bigg( I_\varphi(t_1) I_\varphi(t_2) \left\langle T \bar{\varphi}(t_1, \mathbf{0}) \bar{\varphi}(t_2, \mathbf{0}) \right\rangle \\
&+ I_\varphi^i(t_1) I_\varphi^j(t_2) \left\langle T  \partial_i \bar{\varphi}(t_1, \mathbf{0}) \partial_j \bar{\varphi}(t_2, \mathbf{0}) \right\rangle  +  \frac{1}{4} I_\varphi^{ij}(t_1) I_\varphi^{kl}(t_2) \langle T  \partial_i \partial_j \bar{\varphi}(t_1, \mathbf{0}) \partial_k \partial_l \bar{\varphi}(t_2, \mathbf{0}) \rangle \bigg)\;,
 \end{split}
\ee
where we have included again the symmetry factor of $1/2$ for this diagram. By using the expression of the $\bar{\varphi}$ propagator, eq.~\eqref{eq:propagatorscalar}, and the identities \eqref{eq:ur}, we find
\be
  \hat S_\mathrm{eff}^{(\varphi)} = \frac{1}{2 \Mp^2} \int \frac{d^4 k}{(2 \pi)^4} \frac{1}{k^2 -i \epsilon} \bigg( | I_\varphi (k_0)|^2 + \frac13 |\mathbf{k}|^2   | I_\varphi^i (k_0)|^2 + \frac1{30} |\mathbf{k}|^4   | I_\varphi^{ij} (k_0)|^2  \bigg) \;, 
\ee
where we have introduced the Fourier transforms of the multipole moments, 
\be
I_\varphi(k_0) = \int dt I_\varphi (t) e^{i k_0 t} \;, \quad I_\varphi^i(k_0) = \int dt I_\varphi^i (t) e^{i k_0 t} \;, \quad I_\varphi^{ij}(k_0) = \int dt I_\varphi^{ij} (t) e^{i k_0 t} \;.
\ee

To extract the imaginary part of the above action we use once more eq.~\eqref{eq:ImEpsilon} with \eqref{eq:deltarel}. By an analogous treatment  to that at the end of Sec.~\ref{subsec:radiated_gravitons}, we find the power emitted  into scalars,
\be 
P_\phi = {2G_N}\left[  \big\langle \dot{I}_\varphi^2  \big\rangle  + \frac{1}{3} \big\langle  \ddot{I}_\varphi^i \ddot{I}_\varphi^i  \big\rangle + \frac{1}{30} \big\langle \dddot{I}_\varphi^{ij}  \dddot{I}_\varphi^{ij}   \big\rangle \right] \;.
\label{eq:radiated_power_scalar}
\ee
Therefore, beside the quadrupole,  the monopole and the dipole \cite{Eardley1975ApJ} contribute  as well  to the scalar radiation.

\subsection{Detected signal \label{subsec:detected_signal}}

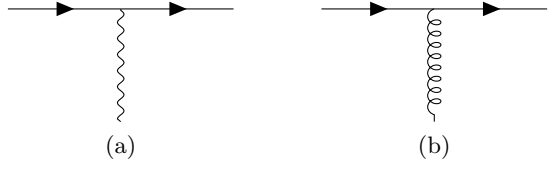
\begin{figure}
	\centering
	\subfloat[]{
		\begin{tikzpicture}
			\begin{feynman}
				\vertex (i1);
				\vertex [right=of i1] (a);
				\vertex [right=of a] (f1);
				\vertex [below=of a] (b);
				
				\diagram*{
				i1 -- [fermion] (a) -- [fermion] (f1),
				(a) -- [photon] (b)
				};
				
			\end{feynman}

		\end{tikzpicture}
		\label{subfig:field_isolated_object_graviton}		
	} \hspace{2em}
		\subfloat[]{
		\begin{tikzpicture}
			\begin{feynman}
				\vertex (i1);
				\vertex [right=of i1] (a);
				\vertex [right=of a] (f1);
				\vertex [below=of a] (b);
				
				\diagram*{
				i1 -- [fermion] (a) -- [fermion] (f1),
				(a) -- [gluon] (b)
				};
				
			\end{feynman}

		\end{tikzpicture}
	              \label{subfig:field_isolated_object_scalar}
} \hspace{1em}

\caption{Feynman diagram giving the radiation field emitted by an object with energy-momentum tensor $T^{\mu \nu}$.}
\label{fig:field_isolated_object}
\end{figure}

Here we compute the  radiation field in gravitons observed at the detector. To simplify the notation we remove the bar over the radiated fields. We need to evaluate the diagram of Fig.~\ref{subfig:field_isolated_object_graviton}---which amounts to find the solution of the equations of motion---but using  a retarded Green's function instead of the Feynman one, so as to enforce the physical nature of the external field. Using the coupling of a radiation graviton to matter directly expanded in multipoles, as found in eq.~\eqref{eq:interaction_graviton_multipole}, in the transverse-traceless gauge this gives \begin{align}
\begin{split}
h_{ij}^{\rm TT}(t, \mathbf{x}) 
= - \frac{i}{\Mp} \Lambda_{ij, kl} \int dt' G_R(t-t', \mathbf{x}) \ddot{I}_h^{kl} (t', \mathbf{0}) \;,
\end{split}
\end{align}
where $ G_R(t-t', \mathbf{x})$ denotes the retarded Green's function between the source located at $(t', \mathbf{0})$ and the observation made at $(t, \mathbf{x})$.
Note that the retarded Green's function is given by a different $i\epsilon$ prescription, which amounts to pick only physical waves modes. In particular, 
\begin{align}
\begin{split}
G_R(t-t', \mathbf{x} - \mathbf{x}') &= \int \frac{d^4k}{(2\pi)^4} \frac{-i}{-(k^0-i\epsilon)^2+\mathbf{k}^2} e^{-ik\cdot (x-x')} \\
&= \frac{i}{4\pi |\mathbf{x}-\mathbf{x}'|} \delta(t'-t-|\mathbf{x}-\mathbf{x}'|) \;.
\end{split}
\end{align}
The second equality comes from the residue theorem. 
Finally, the observed wave (normalized with the Planck mass, so as to agree with the GW literature) is  given by 
\begin{equation}
\frac{h_{ij}^{\rm TT} (t, \mathbf{x})}{\Mp} = \frac{2G_N}{R} \Lambda_{ij,kl} \ddot{I}_h^{kl}(t_\mathrm{ret}) \;,
\label{eq:obsgraviton}
\end{equation}
where $R $ is the distance to the source, $R = | \mathbf{x}|$, and  $t_\mathrm{ret} = t- R$ is the retarded time. 

The scalar waveform can be found by similar reasoning, evaluating  the diagram of Fig.~\ref{subfig:field_isolated_object_scalar} with  the coupling of a radiation field to matter directly expanded in multipoles, as  in eq.~\eqref{eq:multipole_expansion_scalar}.
Given an on-shell scalar wave propagating in the direction $\mathbf{n}$, we can use   $\partial_i \phi = - n_i \partial_t \phi$ and rewrite these couplings as
\begin{equation}
\hat{S}^{(\varphi)}_\mathrm{int} = \frac{1}{\Mp} \int dt \, \bar \varphi \left( I_\varphi + n_i \dot{I}_\varphi^i + \frac{n_i n_j}{2} \ddot{I}_\varphi^{ij}  \right) \;,
\end{equation}
so that the  observed radiation field into scalars reads 
\be
\varphi (t,\mathbf{x}) = \frac{i}{\Mp} \int dt'  \left( I_\varphi (t', \mathbf{0}) + n_i \dot{I}_\varphi^i (t', \mathbf{0}) + \frac{n_i n_j}{2} \ddot{I}_\varphi^{ij}  (t', \mathbf{0}) \right)  G_R(t-t', \mathbf{x})\;. 
\ee
By a treatment analogous to the one for gravitons, we find the radiated field away from the source,
\begin{equation}
\frac{\varphi(t, \mathbf{x})}{\Mp} = - \frac{2G_N}{R} \left( \left. I_\varphi + n_i \dot{I}_\varphi^i + \frac{n_i n_j}{2} \ddot{I}_\varphi^{ij}  \right)\right|_{t_\mathrm{ret}} \;.
\label{eq:obsscalar}
\end{equation}

We can now turn to the effect of the gravitational wave passage on the detector. We denote by $\xi_i$ the separation between two  test masses---for instance the mirrors of a  detector---located at a distance shorter than the typical spatial variation of a gravitational wave. In the proper detector frame, i.e.~choosing coordinates such that the spacetime metric is flat up to tidal effects even during the passage of a gravitational wave, the acceleration between the two masses is given by
 (see e.g.~\cite{Maggiore:1900zz})
\begin{equation}
\ddot{\xi}_i = - R_{i0j0}    \xi_j \;,
\label{eq:motion_riemann}
\end{equation}
where 
\be
R_{i0j0} = \frac{1}{2 \Mp} \left( \partial_i \dot{ {h}}_{0j} + \partial_j \dot{ {h}}_{0i} - \partial_i \partial_j {h}_{00} - \ddot{ {h}}_{ij} \right) \;.
\ee

In standard GR, computing the emitted gravitational wave in the transverse-traceless gauge (so that $h_{00} = h_{0i} = 0$) we have 
\be
R_{i0j0} = -\frac{1}{2 \Mp} \ddot{h}_{ij}^{\rm TT}\;.
\ee 
However, here the detector is non-minimally coupled to $g_{\mu \nu}$. Its scalar charge will generally depend on the local scalar field value (which may be different from the scalar environment of the binary objects) and on the  renormalization effects discessed in Sec.~\ref{sec2.2}.
Defining by $\aaa_{\rm det}$ the  scalar charge of the detector, this can be found by 
\begin{equation}
\int d\tilde{\tau} =  \int d\tau (1-\aaa_{\rm det}\varphi)\;,
\end{equation}
which tells us that, to linear order in the fields,  the physical metric is  
\begin{equation}
\tilde{h}_{\mu \nu} = h_{\mu \nu} - 2\aaa_{\rm det}\varphi \eta_{\mu \nu}\;.
\end{equation}

Using this metric to compute the components of the Riemann tensor in eq.~\eqref{eq:motion_riemann}, we find
\be
\begin{split}
R_{i0j0} &= \frac{1}{2\Mp} \left( \partial_i \dot{ \tilde{h}}_{0j} + \partial_j \dot{ \tilde{h}}_{0i} - \partial_i \partial_j \tilde{h}_{00} - \ddot{ \tilde{h}}_{ij} \right) \\
&= - \frac{1}{2\Mp} \ddot h_{ij}^{\rm TT} + \aaa_{\rm det} \left( \delta_{ij} \ddot \varphi - \partial_i \partial_j \varphi \right)  = - \frac{1}{2\Mp} \partial_t^2 \left[  h_{ij}^{\rm TT} - 2\aaa_{\rm det} \varphi \left( \delta_{ij} - n_i n_j \right)  \right] \;,
\label{eq:physical_GW}
\end{split}
\ee
where in the last equality we have used again $\partial_i \varphi = - n_i \dot \varphi$. Using the expressions for the observed graviton and scalar waves, eqs.~\eqref{eq:obsgraviton} and \eqref{eq:obsscalar}, the detector will observe the following metric perturbation,
\be
\frac{h_{ij}^{\rm detector}}{\Mp}  = \frac{2G_N}{R} \left[  \Lambda_{ij,kl} \ddot I_h^{kl} + 2 \alpha_{\rm det} \left( \delta_{ij} - n_i n_j \right)  \left(  I_\varphi + n_k \dot{I}_\varphi^k + \frac{n_k n_l}{2} \ddot{I}_\varphi^{kl}  \right) \right]_{t_{\rm ret}} \;,
\label{eq:hij_detector}
\ee
For a wave propagating in the direction $ \mathbf{n}$, it is convenient to define the three polarization tensors 
\be
e_{ij}^+  \equiv \mathbf{e}_i \mathbf{e}_j -  \bar {\mathbf{e}}_i \bar {\mathbf{e}}_j \;, \qquad e_{ij}^\times \equiv \mathbf{e}_i \bar {\mathbf{e}}_j + \bar {\mathbf{e}}_i \mathbf{e}_j \;,  \qquad e_{ij}^\phi \equiv \mathbf{e}_i \mathbf{e}_j + \bar {\mathbf{e}}_i \bar {\mathbf{e}}_j  \;, 
\ee
where $\mathbf{e}$ and $ \bar {\mathbf{e}}$ are two unit vectors defining an orthonormal basis with $ \mathbf{n}$.
We can then decompose the metric into these three polarization states, 
\be
\frac{h_{ij}^{\rm detector}}{\Mp}  = \sum_{s= +, \times, \phi} e^s_{ij}( \mathbf{n})  h_s \;,
\ee
where for the two standard transverse-traceless polarizations we have
\be
\label{eq:hplushcross}
h_{+, \times} = \frac{G_N}{R}  e_{ij}^{+, \times} ( \mathbf{n})  \ddot I_{h}^{ij}( {t_{\rm ret}}) \;,
\ee
while for the additional scalar polarisation we find
\be
h_\phi =  \frac{4 \alpha_{\rm det} G_N}{R}  \left(  I_\varphi + n_k \dot{I}_\varphi^k + \frac{n_k n_l}{2} \ddot{I}_\varphi^{kl}  \right) \Big|_{t_{\rm ret}}  \;.
\label{eq:scalar_polarization}
\ee

\subsection{Circular orbits}

We  now  compute  the wave amplitudes emitted by two binary objects in terms of the binary system parameters.
As before, we limit our calculation to the lowest post-Newtonian order. 

As the emission of GW circularizes the orbit, we assume a circular orbit in which the relative coordinate of the system, $\mathbf{r}= \mathbf{x}_1 - \mathbf{x}_2$, has cartesian components parametrized in time as 
\begin{align}
\begin{split}
r_x(t) &= r \cos(\omega t + \pi/2) \;, \\
r_y(t) &= r \sin(\omega t + \pi/2) \;, \\
r_z(t) &= 0 \;.
\end{split}
\end{align}
We  first assume that the frequency of the binary $\omega $ is constant. In the next subsection we will consider its time dependence due to the backreaction of the GW emission on the circular motion. For the following discussion it is convenient to define  the reduced mass of the system $\mu$ and the total mass $M$ as
\be
\mu \equiv \frac{m_1m_2}{M}  \;, \qquad M \equiv m_1+m_2 \;.
\ee

We chose the axis of rotation of the binary system to coincide with the $\mathbf{z}$ axis  while the propagation vector of the GW is  oriented in an arbitrary direction parametrized by the angles $\theta$ and $\phi$, 
\begin{equation}
\mathbf{n} \equiv (\sin \theta \sin \phi, \sin \theta \cos \phi, \cos \theta) \;.
\end{equation}
For the gravitational polarizations $h_+$ and $h_\times$, replacing the above expressions in the quadrupole moment given by eq.~\eqref{eq:quadrupole_gr}, and using this in eq.~\eqref{eq:hplushcross}, one finds (see e.g.~\cite{Maggiore:1900zz})
\begin{align}
\begin{split}
h_+ &= \frac{4 G_N \mu (\omega r)^2}{R} \left( \frac{1+\cos^2\theta}{2} \right) \cos(2\omega t_\mathrm{ret} +2\phi)\;,\\
h_\times &= \frac{4 G_N \mu (\omega r)^2}{R} \cos \theta \sin(2\omega t_\mathrm{ret} +2\phi) \;.
\end{split}
\end{align}
By using Kepler's third law to lowest order, i.e.
\be
\label{eq:Kepler}
\omega^2 = \frac{\tilde{G}_{12}M}{r^3} 
\ee
(we remind that $\tilde G_{12} = (1+2 \aaa_1 \aaa_2) G_N$, see eq.~\eqref{eq:Gab}), we find $\omega r = ( \tilde{G}_{12}M \omega)^{1/3}$.  
Note that this quantity scales as $v$. Using this expression in eq.~\eqref{eq:hgrav} to eliminate $r$,  we can rewrite the scalar waveform as 
\begin{align}
\label{eq:hgrav}
\begin{split}
h_+ &= \frac{4 G_N \mu}{R} (\tilde{G}_{12}M\omega)^{2/3} \left( \frac{1+\cos^2\theta}{2} \right) \cos(2\omega t_\mathrm{ret} +2\phi) \;, \\
h_\times &= \frac{4 G_N \mu}{R} (\tilde{G}_{12}M\omega)^{2/3} \cos \theta \sin(2\omega t_\mathrm{ret} +2\phi) \;.
\end{split}
\end{align}

Let us now turn to the scalar polarization, given by eq.~\eqref{eq:scalar_polarization}. 
For circular motion  the monopole term is constant in time and can be discarded. Using the center-of-mass relation~\eqref{eq:center_of_mass} to compute the time derivative of the dipole, $n_k \dot{I}_\varphi^k$ in eq.~\eqref{eq:scalar_dipole}, 
and eliminating the $r$ dependence using eq.~\eqref{eq:Kepler} above, we find the dipolar scalar emission to lowest order,
\begin{align}
\label{dipoleee}
h^{\rm dipole}_\phi = - \frac{4\aaa_{\rm det} G_N \mu}{R}  (\aaa_1 - \aaa_2) (\tilde{G}_{12}M\omega)^{1/3} \sin \theta \sin(\omega t_\mathrm{ret}+\phi)  \;.
\end{align}
 Similarly, we can compute the second time derivative of the quadrupole moment, ${n_k n_l} \ddot{I}_\varphi^{kl}$  in eq.~\eqref{eq:scalar_dipole}, and find the quadrupolar scalar emission,
\begin{align}
\label{quadrupoleee}
h^{\rm quadrupole}_\phi = - \frac{4\aaa_{\rm det} G_N \mu}{R} \frac{\aaa_1 m_2 +\aaa_2 m_1}{M}  (\tilde{G}_{12}M\omega)^{2/3} \sin^2 \theta \cos(2\omega t_\mathrm{ret}+2\phi)   \;.
\end{align}

Few comments are in order here. First, notice that  $\aaa_{\rm det}$ is the coupling of the detector to the scalar, while the $\aaa_A$'s are the renormalized couplings of the inspiral objects, which can depend on their masses. Since $\alpha \ll 1$, we expect the scalar amplitude of the GW to be suppressed with respect to the gravitational one.
Second, comparing the powers of the combination $(\tilde{G}_{12}M\omega)^{1/3} \sim v$ in eq.~\eqref{dipoleee} and in eqs.~\eqref{eq:hgrav} and \eqref{quadrupoleee}   confirms that the dipole is of 0.5PN order less than the gravitational quadrupole, as expected.

\subsection{Frequency dependence}
\label{subsec:frequency_dependance}
Because the number of gravitational wave oscillations within a typical LIGO/Virgo event is very large, gravitational wave detectors are much more sensitive to a phase change rather than a modification of the amplitude. 
For this reason, in this subsection we will compute the frequency dependence of the waveform  from our formalism.

To this aim, we can use the energy balance of the system, i.e.~that the  total power loss is equal to the time derivative of the orbital energy. This reads
\be
\label{energy_balance}
P_g+P_\phi^\mathrm{monopole}+ P_\phi^\mathrm{dipole} + P_\phi^\mathrm{quadrupole} =  - \frac{dE}{dt} \;,
\ee
where $P_g$, $P_\phi^\mathrm{monopole}$, $P_\phi^\mathrm{dipole}$ and $P_\phi^\mathrm{quadrupole}$ are respectively the graviton, and the scalar monopole,  dipole and quadrupole  contributions to the emitted power.
As explained above, since the scalar monopole is constant for circular orbits, its emitted power   vanishes, $P_\phi^\mathrm{monopole}=0$. The orbital energy is given by
\be
E\equiv -\frac{\tilde{G}_{12}m_1m_2}{2r} = - \frac12 (\tilde{G}_{12} M_c \omega )^{2/3} M_c\;,
\ee
where for the last equality we have used again the Kepler's law and we have defined the chirp mass,
\be
M_c \equiv \frac{(m_1 m_2)^{3/5}}{M^{1/5}}= \mu^{3/5} M^{2/5}\;.
\ee

From eq.~\eqref{eq:quadrupole_graviton}, the power emitted  into gravitons reads
\begin{equation}
P_g = \frac{32}{5} G_N \mu^2 \omega^6 r^4 \;
\end{equation}
and, using again Kepler's law, one can rewrite this  as 
\begin{equation}
P_g = \frac{32}{5\tilde{G}_{12} (1+2\aaa_1 \aaa_2)} (\tilde{G}_{12} M_c \omega)^{10/3} \;.
\end{equation}

One can then proceed analogously for the power emitted into scalars. At lowest order in $v$, the power emitted by the scalar dipole contribution reads
\begin{equation} \label{eq:power_dipole}
P_\phi^\mathrm{dipole} = \frac{2}{3\tilde{G}_{12} (1+2\aaa_1 \aaa_2)} (\aaa_1-\aaa_2)^2 \nu^{2/5} (\tilde{G}_{12} M_c \omega)^{8/3} \;,
\end{equation}
where 
\be
\nu \equiv \frac{m_1m_2}{M^2}
\ee 
is the symmetric mass ratio.

We have derived eq.~\eqref{eq:power_dipole} at lowest order in the velocity expansion. But the quadrupolar power is suppressed by $v^2$ compared to the dipolar one (the Feynman diagrams of Fig.~\ref{fig:radiation_Seff} giving the radiated power involve two interaction vertices that are respectively of order $\sqrt{Lv^3}$ for the dipole and $\sqrt{Lv^5}$ for the quadrupole). We need therefore to compute the dipolar power at next-to-leading order so as to find an expression consistent with the quadrupolar order. 
To simplify the discussion, 
we  discard this correction  here. This approximation can then be used when $\aaa_1-\aaa_2 \lesssim v$, so that the dipole is smaller or of the same order as the quadrupole and its $v^2$ corrections are thus negligible or, alternatively, when $\aaa_1-\aaa_2 \gg v$, in which case the dipole dominates and we can ignore the quadrupolar terms. 
For completeness, we compute the dipolar power at next-to-leading in App. \ref{appB}.

Finally, from eq.~\eqref{eq:radiated_power_scalar} one finds that the power emitted by the scalar quadrupole contribution is proportional to that of the gravitational   quadrupole, i.e.,
\begin{equation}
P_\phi^\mathrm{quadrupole} =   \frac{\left( \aaa_1 m_2 +\aaa_2 m_1\right)^2 }{3 M^2}  P_g \;.
\end{equation}

Using these expressions into the left-hand side of the energy balance equation, eq.~\eqref{energy_balance}, we can find a differential equation for the time derivative of the frequency. Following \cite{Will:1994fb}, it is convenient to define the scalar-tensor chirp mass, 
\be
\tilde M_c^{5/3} \equiv \frac{M_c^{5/3}}{1+2\aaa_1 \aaa_2} \left[1+\frac{(\aaa_1 m_2 + \aaa_2 m_1)^2}{3M^2} \right] \; , 
\ee
and the dipole parameter,
\be
b \equiv \frac{5}{48} (\aaa_1-\aaa_2)^2 \frac{M_c}{\tilde M_c}  \; .
\ee
In terms of these quantities, the evolution equation for $\omega$ reads
\begin{equation} \label{eq:evolution_omega_rescaled}
\dot \omega = \frac{96}{5} (\tilde G_{12} \tilde M_c)^{5/3} \omega^{11/3} \left[1+b \nu^{2/5} (\tilde G_{12} \tilde M_c \omega)^{-2/3} \right] \; .
\end{equation}

We compute the total phase accumulated in the GW detector,  focussing on the quadrupole.
This reads
\begin{equation}
\Phi_{\rm quadrupole} = 2 \int dt \, \omega(t) = 2 \int d\omega \frac{\omega}{\dot \omega} \; ,
\end{equation}
where the factor of two comes from the frequency dependence of the quadrupolar waveform \eqref{eq:hgrav}. Expanding for small $b (\tilde G_{12} \tilde M_c \omega)^{-2/3} \ll 1$, we can integrate eq.~\eqref{eq:evolution_omega_rescaled} to get
\begin{equation}
\Phi_{\rm quadrupole} = \frac{1}{16} \left[ (\tilde G_{12} \tilde M_c  \pi f)^{-5/3}  - \frac{5}{7} b \nu^{2/5} (\tilde G_{12} \tilde M_c  \pi f)^{-7/3} \right]_{f_\mathrm{in}}^{f_\mathrm{out}} \; ,
\end{equation}
where we used $f \equiv  \omega/ \pi$ to convert the angular frequency of the binary system into the GW frequency emitted by the quadrupole \footnote{Note that there could be a conformal rescaling of the frequency from the time of the GW emission to the one of its detection, due to the cosmological evolution of the field. See the end of Sec.~6 of \cite{damour_tensor-multi-scalar_1992}. }.
Moreover, $f_\mathrm{in}$ ($f_\mathrm{out}$) denotes the frequency at which the GW signal enters (exits) the  detector. For LIGO/Virgo, we have $f_\mathrm{in} \sim 10\,$Hz $\ll f_\mathrm{out} \sim 1\,$kHz. By requiring that the phase modification is less than $\pi$, we obtain the following approximate bound on the dipole parameter $b$,
\begin{equation}
b \lesssim \frac{112 \pi}{5} \nu^{-2/5} (\tilde G_{12} \tilde M_c \pi f_\mathrm{in})^{7/3}  \simeq 10^{-6}\; .
\end{equation}
Note that the strongest constraint comes from the beginning of the inspiral, when the signal at $f_\mathrm{in} \sim 10$ Hz enters the detector. Our results are in agreement with earlier work by Will \cite{Will:1994fb}, which uses the sensitivities $s_A$ defined in eq.~\eqref{eq:sensitivity}, instead of the parameters $\alpha_A$. Moreover, the  waveform in scalar-tensor gravity has been computed up to 2PN order in \cite{Sennett:2016klh}.

\section{Extensions}\label{Sec:dis}

As discussed in the introduction,  dark energy models generally feature non-linearities that become important in the vicinity of a massive body and can screen the effect of the scalar field. When present, such non-linearities  make our diagrammatic expansion meaningless. In the Feynman perturbative expansion, propagators represent the free part of the Lagrangian, which dominates the dynamics, while interactions are treated perturbatively.
This is no longer the case close to the source.
 
However, we can consider an extension to the models studied in this paper  in which we can trust our usual propagator and where non-linearities show up in a more subtle way. 
Consider the {\it disformal coupling} of eq.~\eqref{disformal}, relating the Jordan  frame metric $\tilde{g}_{\mu \nu}$ to the Einstein frame one $g_{\mu \nu}$.
A similar disformal coupling has been studied in theories of dark energy (see e.g.~\cite{Gleyzes:2015pma}), where its natural value  is $\Lambda_* \sim \Lambda_2 \equiv (H_0 \Mp)^{1/2}$.
In the Einstein frame, the standard point-particle minimal coupling to $\tilde{g}_{\mu \nu}$ induces, on top of other various terms included in our point-particle action \eqref{eq:pp_Action_expanded}, a vertex of the type
\begin{equation}
\frac{m_A}{\Lambda_*^4} \int dt (\partial_\mu \varphi v_A^\mu)^2\, ,
\label{eq:disformal_vertex}
\end{equation}
which is part of the more general point-particle action discussed in the App.~\ref{app}. Note that, for simplicity, we are assuming this coupling to be universal, i.e.~dependent only on the mass $m_A$ of the object and not on other object-dependent couplings.

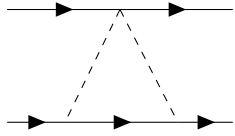
\begin{figure}
	\centering
		\begin{tikzpicture}
			\begin{feynman}
				\vertex (i1);
				\vertex [right=of i1] (a);
				\vertex [right=of a] (f1);
				\vertex [below=of i1] (i2);
				\vertex [right=2em of i2] (b);
				\vertex [right=of b] (c);
				\vertex [below=of f1] (f2);
				
				\diagram*{
				(i1) -- [fermion] (a) -- [fermion] (f1),
				(i2) -- [fermion] (b) -- [fermion] (c) -- [fermion] (f2),
				(a) -- [scalar] {(b), (c)},
				};
				
			\end{feynman}
		
		\end{tikzpicture}
\caption{Feynman diagram corresponding to the first disformal correction to the conservative dynamics (it should also be added with its symmetric counterpart). The upper vertex is the one of eq. \eqref{eq:disformal_vertex}.}
\label{fig:disformal}
\end{figure}

The first contribution of this new vertex to the conservative dynamics of the objects is given by the diagram of Fig.~\ref{fig:disformal}, and can be calculated to be
\begin{equation}
\mathrm{Fig}. \ \ref{fig:disformal} =  i \frac{G_N m_1 m_2 (\aaa_1^2 m_1 + \aaa_2^2 m_2)}{2 \pi \Lambda_*^4} \int dt \frac{({\mathbf{r}} \cdot \mathbf{v})^2}{r^6}\, ,
\label{eq:disformal_energy}
\end{equation}
where $\mathbf{v} =  \mathbf{v}_1 - \mathbf{v}_2$.

It is interesting to compare this quantity with the simple Newtonian potential ${G_N m_1 m_2}/{r}$ plotted in Fig.~\ref{fig:Newt_pot_A}. Recalling that the virial theorem gives the approximate relation $v^2 \sim {G_N m}/{r}$, we get the following estimate,
\begin{equation}
\frac{\mathrm{Fig.}\  \ref{fig:disformal} }{ \mathrm{Fig.} \ \ref{fig:Newt_pot_A} }\ \sim \ \left(  \frac{ \aaa m}{\Mp \Lambda_*^2 r^2} \right)^2 \ \sim \ \left( \frac{r_*}{r} \right)^4\;,
\end{equation}
where we have assumed that  scalar couplings and masses are roughly the same, $\aaa_1\sim \aaa_2 \sim \alpha$ and $ m_1\sim m_2\sim m$, and in the last equality we have  introduced the non-linear radius,\footnote{
In the $k$-mouflage screening mechanism~\cite{Babichev:2009ee}, the  scalar field Lagrangian contains a quadratic term and a non-linear term suppressed by a strong coupling scale $\Lambda_*$, such as in eq.~\eqref{eq:kmoufla}.
The lengthscale $r_*$ defined in eq.~\eqref{eq:Vradius} is exactly the  radius  at which the non-linear term above  dominates over the quadratic one around overdense sources and inside which the fifth force is screened.} 
\be
\label{eq:Vradius}
r_* = \frac{1}{\Lambda_*} \left( \frac{\alpha m}{\Mp}   \right)^{1/2}\;.
\ee
For $r \lesssim r_*$ the new term dominates  on the  Newtonian interaction, signalling a breakdown of the diagrammatic expansion.  In order for our perturbative calculations to be predictive we should tune the mass parameter $\Lambda_*$ in such a way that $r_*$ is smaller than the size $r$ of the system. This would correspond to dealing with another independent UV scale on top of the Schwarzschild radius $r_s \ll r$: $r_*\ll r$. 
From eq.~\eqref{eq:Vradius} above one finds, for a system of the mass of the Sun,
\be
\Lambda_* \sim \alpha^{1/2}  \frac{r_s}{r_*} 10^{11} \Lambda_2  \;,
\ee
which for $r_* < r_s$ and sizeable $\alpha$ is  much larger than the typically expected value $\Lambda_2$.

Let us stress the difference between such a non-linear coupling and those displayed by $k$-mouflage and Galileon-like theories in screened regions. The latter contain non-linear terms in the evolution equation of the scalar field, that show up directly in the spherically symmetric solution of the scalar field configuration generated by a static source. Equivalently, these terms, which become leading close to the source, do not allow to use the standard propagator in a diagrammatic expansion. Here, non linearities are all  hidden in the  coupling to the point-particle. In the  vacuum the field obeys the usual Laplace equation, $\nabla^2 \varphi = 0$, and does not exhibit any transition to a Vainshtein regime at small radii. The standard spherically symmetric/static analysis (equivalently, the one-body diagrams like those, say, of Fig.~\ref{fig:field_isolated_object}) cannot grasp the non-linear dynamical aspects of the disformal model, which are encoded in the velocity dependent {\it two} body diagrams like the one if Fig.~\ref{fig:disformal}.  The effects of such a disformal coupling on the dynamics of gravitationally interacting bodies, already initiated in \cite{Brax:2018bow}, will be further explored in a future publication~\cite{Inprep}

\section{Concluding remarks}

In this paper we have generalized to scalar-tensor theories the EFT formalism of Goldberger and Rothstein~\cite{goldberger_effective_2006} for gravitational wave emission from a binary system. With an eye on dark energy, we have assumed the scalar field to be massless. The basic power counting of the Feynman diagrams in the relative velocity $v$ between the two objects follows quite closely that of pure gravity. 
For most of the paper we have considered standard conformal couplings of the scalar to point particles. We have discussed violations of the strong equivalence principle in terms of the matching conditions 
between the UV model of the body (the one that ``knows" about its size, density distribution, etc.) and the (low energy) point-particle description. 
By integrating out potential gravitons and scalars we recover the extended EIH Lagrangian that corrects the conservative Newtonian dynamics to relative order $v^2$. Our results are consistent with those of Damour and Esposito Far\`{e}se~\cite{damour_tensor-multi-scalar_1992}. Finally, we have obtained the radiated power in gravitational and scalar waves by integrating out the {\it radiation} gravitons and scalars. At the same time, by using the retarded propagator, we have worked out the waveform in the presence of a scalar field.

The latter has several potential distinct features~\cite{Will:2014kxa}. 
The additional power loss in scalar radiation (monopole, dipole and quadrupole) modifies the dynamics of the system,  and so the time evolution of the frequency of the GW,  which ultimately modifies its phase.  Moreover, the dipole radiation, proportional to the {\it difference} of the scalar charges of the two bodies, has the same frequency  $\omega$ as the binary, as opposed to the quadrupole radiation that has frequency $2\omega$. Finally, in the presence of a scalar field, there will be an additional polarization associated to GW (see eq.~\eqref{eq:physical_GW}). It should be noted that these last two effects modify the amplitude of the GW signal, which is far less constrained than the phase by detectors.

The model that we have considered is an important test bench for generalizing the EFT formalism of~\cite{goldberger_effective_2006} to modified gravity, but it also contains obvious limits. Perhaps the most serious one is that its observational signatures will be very hard to detect. The departures from GR that we have just mentioned are proportional to the scalar coupling $\alpha$, which however is constrained to be less than $10^{-2}$ by Solar System tests~\cite{Bertotti:2003rm}. More realistic models of dark energy, on the other hand, contain non-linearities in the scalar dynamics that are difficult to deal with. As discussed in Sec.~\ref{Sec:dis}, an interesting non-linear behavior emerges dynamically in models with standard kinetic terms for the graviton and the scalar and disformal couplings to the point particle. Other non-linearities of the $k$-essence or Galileon-type are the subject of ongoing and future work.

\section*{Acknowledgments} We thank Philippe Brax, Matt Johnson, Alberto Nicolis and Massimiliano Riva for useful discussions. 

\appendix

\section{General couplings to point particles}\label{app}

Cosmological models of dark energy and modified gravity with a single scalar field are very conveniently studied in the unitary gauge, where the time coordinates are set in such a way to coincide with the uniform-field hypersurfaces  (see e.g.~\cite{Piazza:2013coa} for more details). This allows, with relative ease, to write down the most general action for the gravity/scalar sector that is consistent with the symmetries that are usually unbroken in a cosmological set up: spatial translations and rotations~\cite{Creminelli:2006xe,Cheung:2007st,Gubitosi:2012hu,Bloomfield:2012ff}. In the unitary gauge all degrees of freedom end up encoded in the metric field. The presence of the scalar manifests itself in all those Lagrangian terms that (spontaneously) break {\it time-diffeomorphisms}. General covariance can then be recovered with the St\"uckelberg procedure. 

Before turning to the covariant formalism, it is interesting to look at the possible matter couplings in the unitary gauge.
We are interested in a point-like source described by a trajectory $x^\mu = x^\mu(\lambda)$, whose action is invariant under 
\begin{itemize}
\item[$i)$] worldline reparametrization $\lambda \rightarrow {\tilde \lambda}(\lambda)$, 
\item[$ii)$] spatial rotations $SO(3)$ in the Lorentz frame of the object, 
\item[$iii)$] spatial diffeomorphisms but not necessarily {\it time}-diffeomorphisms. 
\end{itemize}
As described in~\cite{goldberger_effective_2006}, worldline reparameterization is taken care of simply by using  the proper time variable $d\tau = d \lambda \sqrt{g_{\mu \nu} \frac{d  x^\mu}{d \lambda}\frac{d x^\nu}{d\lambda}}$ as the worldline parameter. $SO(3)$ invariance restricts the analysis to spherically symmetric spin-less objects. On the other hand, relaxing invariance under time-diffs allows for new terms in the particle Lagrangian. In particular,
\begin{enumerate}
\item All couplings  can be explicitly time-dependent. To lowest order in the $v$ expansion, this simply applies to the mass parameter,
\begin{equation} \label{masss}
\underset{\rm GR}{\int d\tau \ m} \quad \longrightarrow \quad \underset{\rm GR +  scalar}{\int d\tau \ m(t)}\, .
\end{equation}
\item The point particles can couple directly to metric invariants (four-dimensional scalars) multiplied by functions of the time. In a derivative expansion, after the mass term~\eqref{masss} we have
\begin{equation}
\int d \tau \, c_R(t) R\ + \ \int d\tau \, c_V(t) R_{\mu \nu} v^\mu v^\nu\, .
\end{equation}
It is easy to see that, like in the standard case (see the details in~\cite{Goldberger:2007hy}), these terms can be redefined away with a metric field redefinition because they are proportional to the equations of motions in the vacuum. The first non-trivial of such couplings involve two powers of the (Weyl) curvature and contribute only to order $v^{10}$ to the two body Lagrangian~\cite{goldberger_effective_2006}. 
\item The point particles can couple also to {\it three-dimensional} scalars, i.e.~quantities that are scalars from the point of view of spatial diffeomorphisms, such as $g^{00}$,  
\begin{equation}
S_{\rm pp} \supset  \int d\tau \  g^{00} \mu(t)\, . 
\end{equation}
\item The $t=$ const. hypersurfaces, up to a spatial rotation, pick out a preferred Lorentz frame at any point-event. This can be represented by the unitary vector $n_\mu= \frac{\delta_\mu^0}{\sqrt{-g^{00}}}$ orthogonal to this hypersurface. The (squared) velocity of the point particle with respect to such a frame  
\begin{equation}\label{V}
V^2 = 1 - \frac{1}{(n_\mu v^\mu)^2}\, 
\end{equation}
is an invariant which can appear in the point-particle Lagrangian.\footnote{We can indeed introduce a Lorentzian tetrad $e^\mu_{(\alpha)}$ at any point and define $V^2$  through the {\it boost} transformation that must be made upon  $n^{(\alpha)}$ in order to obtain $v^{(\alpha)}$. $n_\mu v^\mu = n_{(\alpha)} v^{(\alpha)}$ is just the gamma factor of such a boost. On the other hand, it is customary to define $v^i$  as the {\it coordinate} velocity of the particle, $v^i = \frac{d x^i}{d t}$.}
\end{enumerate}

In summary, in the presence of a scalar degree of freedom we can expect a particle Lagrangian of the type
\begin{equation}\label{spp}
S_{\rm pp} = \int d\tau \left[m(t) + \mu(t) \delta g^{00} + \bar m(t) V^2 + \dots \right]\, .
\end{equation}
In the above, we have used $\delta g^{00} = 1 + g^{00}$ to avoid redundancies with the mass term. 
In order to make the couplings with the scalar explicit, we force a time diffeomorphism (St\"uckelberg procedure) $t\rightarrow t+\pi(x)$, and interpret $\pi$ as the scalar field perturbation. It is immediate to recover the standard scalar-point-particle couplings of the ``Brans Dicke type" in the Einstein frame,
\begin{equation}
m(t) \ \rightarrow \ m(t+\pi)\ = \ m(t) + \dot m(t) \pi(x) + \frac12 \ddot m(t) \pi^2(x) + \dots\, .
\end{equation} 
The term $g^{00}$, on the other hand, produces derivative couplings. By using the standard transformation property of the metric under a diffeomorphism one gets
\begin{equation}
g^{00}\ \rightarrow \ g^{00} + 2 g^{0\mu}\partial_\mu \pi + g^{\mu \nu}\partial_\mu \pi \partial_\nu \pi\, .
\end{equation}
Finally, $V^2$ in~\eqref{V} contains a rather rich structure, 
\begin{equation}
V^2 \ = \ 1+ \frac{g^{00}}{\left(\partial_\mu t \  \dfrac{d x^\mu}{d\tau}\right)^2} \ \rightarrow \ 1+ \frac{g^{00} + 2 g^{0\mu} \partial_\mu \pi + g^{\mu \nu}\partial_\mu \pi \partial_\nu \pi}{(1\ +\ \dot\pi + \partial_i \pi v^i)^2} \ \left(\frac{d\tau}{dt}\right)^2\, .
\end{equation}

Action~\eqref{spp} becomes
\begin{align}\label{line13}
S_{\rm pp} =& \int d\tau \left\{m(t) + \dot m(t) \pi(x)  + \frac12 \ddot m(t) \pi^2+ \dots \right.\\
+ &\left[\mu(t) + \dot \mu(t) \pi(x) + \frac12 \ddot \mu(t) \pi^2+ \dots \right]\cdot \left[\delta g^{00} - 2 \dot \pi + 2\,  \delta g^{0i}\, \partial_i\pi +\dots  \right] \\
+ &\left. \left[\bar m(t) + \dot {\bar m}(t) \pi(x) + \frac12 \ddot {\bar m}(t) \pi^2+ \dots \right]\cdot \left[v^2 + 2 v^i \partial_i\pi + v^i \delta g_{0i} +\dots  \right] + \dots \right\} \;. \label{line15}
\end{align}
In the above expression terms have been ordered also according to their power counting in $v$ (see discussion in Sec.~\ref{sec3.3}). Of course, we are in the presence of a much richer set of possibilities than those represented by the coupling metric-matter, usually excluded by the equivalence principle.

\section{Dipolar dissipated power at next-to-leading order} \label{appB}

In order to have  the dipolar radiated power at the same order as the quadrupolar one, we  derive here the first-order correction to the dipolar power. To this aim, we will use the full next-to-leading order dipole in eq. \eqref{eq:dipole_full}. However, we must  correct the center-of-mass relations  in eq.~\eqref{eq:center_of_mass}  by terms higher order in the velocity. 
In particular, using the centre-of-mass definition in eq.~\eqref{eq:def_center_mass} with the $00$-component of the stress-energy tensor given by
\begin{equation}
T^{00} = \sum_A m_A \left(1+\frac{v_A^2}{2} - \frac{\tilde{G}_{12} m_{\bar{A}}}{2r} \right) \delta^3(\mathbf{x} - \mathbf{x}_A ) \; ,
\end{equation}
the center-of-mass relations become
\begin{equation} \label{eq:CM_NLO}
\mathbf{x}_1 = \frac{\mu_2}{\mu_1+\mu_2} \mathbf{r}, \qquad \mathbf{x}_2 = - \frac{\mu_1}{\mu_1+\mu_2} \mathbf{r} \; ,
\end{equation}
with
\begin{equation}
\mu_A \equiv m_A \left(1+\frac{v_A^2}{2} - \frac{\tilde{G}_{12} m_{\bar{A}}}{2r} \right) \; .
\end{equation}

The generalized equations of motion can be found by varying the EIH lagrangian \eqref{eq:LEIH}. In the case of a circular motion,
\begin{equation}
v^2 \equiv (\mathbf{v}_1 - \mathbf{v}_2)^2 = \frac{\tilde{G}_{12}(m_1+m_2)}{r} \; ,
\end{equation}
and using the centre-of-mass relations  above we find, for the acceleration up to next-to-leading order,
\begin{align} \label{eq:EOM_NLO}
\begin{split}
\frac{d^2 \mathbf{r}}{dt^2} &= - \frac{\tilde{G}_{12}M}{r^3} \mathbf{r} \bigg\{1+\frac{\tilde{G}_{12}M}{r} \left[ \nu - \frac{3+2 \aaa_1 \aaa_2}{1+2 \aaa_1 \aaa_2} \right.  \\
& + 4  \left. \frac{ \aaa_1^2 \aaa_2^2 + \beta_1 \aaa^2_2(1+\kappa_{12}) + \beta_2 \aaa_1^2(1-\kappa_{12})}{(1+2 \aaa_1 \aaa_2)^2} \right] \bigg\} \; ,
\end{split}
\end{align}
where we recall  the notation: $M \equiv m_1 + m_2$,  $\nu \equiv \frac{m_1 m_2}{M^2}$ and $ \kappa_{12} \equiv \frac{m_1-m_2}{M}$.

The scalar radiated power by the dipole is given by (see eq.~\eqref{eq:radiated_power_scalar})
\begin{equation}
P_\phi^\mathrm{dipole} = \frac{2G_N}{3} \big\langle  \ddot{I}_\varphi^i \ddot{I}_\varphi^i  \big\rangle \; .
\end{equation}
Inserting the dipole expression up to next-to-leading order in eq.~\eqref{eq:dipole_full} and using eqs.~\eqref{eq:CM_NLO} and \eqref{eq:EOM_NLO}  derived above we finally find, for the dipolar power up to next-to-leading order,
\begin{align}
\begin{split}
P_\phi^\mathrm{dipole} &= \frac{2  (\aaa_1-\aaa_2)^2}{3\tilde{G}_{12} (1+2\aaa_1 \aaa_2)} \nu^{2/5} (\tilde{G}_{12} M_c \omega)^{8/3} \\
& + \frac{4 (\aaa_1-\aaa_2)}{15 \tilde{G}_{12} (1+2\aaa_1 \aaa_2)^3} (\tilde{G}_{12} M_c \omega)^{10/3}  
 \left[17 (\aaa_1-\aaa_2) + (\aaa_1-\aaa_2) \aaa_1 \aaa_2 (2 \aaa_1 \aaa_2-43) \right. \\
&+ \kappa_{12} (\aaa_1+\aaa_2)(1+2\aaa_1 \aaa_2)(4+3\aaa_1 \aaa_2) -2 \kappa_{12}^2 (\aaa_1-\aaa_2) (1+2\aaa_1 \aaa_2)^2 \\
& - \left. 10 \beta_2 \aaa_1(1+\kappa_{12})(1-2\aaa_1(\aaa_1-2\aaa_2)) + 10 \beta_1 \aaa_2(1-\kappa_{12})(1-2\aaa_2(\aaa_2-2\aaa_1))  \right] \; .
\end{split}
\end{align}
After some algebra, it is possible to show that this result agrees with eq. (6.44) of Ref.~\cite{damour_tensor-multi-scalar_1992}. On the other hand, it differs from eq.~(55) of Ref.~\cite{Huang:2018pbu}.

\bibliographystyle{utphys}
\bibliography{bib}

\providecommand{\href}[2]{#2}\begingroup\raggedright\begin{thebibliography}{10}

\bibitem{Abbott:2016blz}
{\bf LIGO Scientific, Virgo} Collaboration, B.~P. Abbott {\em et.~al.},
  ``{Observation of Gravitational Waves from a Binary Black Hole Merger},''
  {\em Phys. Rev. Lett.} {\bf 116} (2016), no.~6 061102,
  \href{http://xxx.lanl.gov/abs/1602.03837}{{\tt 1602.03837}}.

\bibitem{Yunes:2016jcc}
N.~Yunes, K.~Yagi, and F.~Pretorius, ``{Theoretical Physics Implications of the
  Binary Black-Hole Mergers GW150914 and GW151226},'' {\em Phys. Rev.} {\bf
  D94} (2016), no.~8 084002, \href{http://xxx.lanl.gov/abs/1603.08955}{{\tt
  1603.08955}}.

\bibitem{Gleyzes:2013ooa}
J.~Gleyzes, D.~Langlois, F.~Piazza, and F.~Vernizzi, ``{Essential Building
  Blocks of Dark Energy},'' {\em JCAP} {\bf 1308} (2013) 025,
  \href{http://xxx.lanl.gov/abs/1304.4840}{{\tt 1304.4840}}.

\bibitem{Jimenez:2015bwa}
J.~Beltran~Jimenez, F.~Piazza, and H.~Velten, ``{Evading the Vainshtein
  Mechanism with Anomalous Gravitational Wave Speed: Constraints on Modified
  Gravity from Binary Pulsars},'' {\em Phys. Rev. Lett.} {\bf 116} (2016),
  no.~6 061101, \href{http://xxx.lanl.gov/abs/1507.05047}{{\tt 1507.05047}}.

\bibitem{Lombriser:2015sxa}
L.~Lombriser and A.~Taylor, ``{Breaking a Dark Degeneracy with Gravitational
  Waves},'' {\em JCAP} {\bf 1603} (2016), no.~03 031,
  \href{http://xxx.lanl.gov/abs/1509.08458}{{\tt 1509.08458}}.

\bibitem{Bettoni:2016mij}
D.~Bettoni, J.~M. Ezquiaga, K.~Hinterbichler, and M.~Zumalacrregui, ``{Speed of
  Gravitational Waves and the Fate of Scalar-Tensor Gravity},'' {\em Phys.
  Rev.} {\bf D95} (2017), no.~8 084029,
  \href{http://xxx.lanl.gov/abs/1608.01982}{{\tt 1608.01982}}.

\bibitem{the_ligo_scientific_collaboration_gw170817:_2017}
T.~L.~S. Collaboration and T.~V. Collaboration, ``{GW}170817: {Observation} of
  {Gravitational} {Waves} from a {Binary} {Neutron} {Star} {Inspiral},'' {\em
  Physical Review Letters} {\bf 119} (Oct., 2017). arXiv: 1710.05832.

\bibitem{Creminelli:2017sry}
P.~Creminelli and F.~Vernizzi, ``{Dark Energy after GW170817 and GRB170817A},''
  {\em Phys. Rev. Lett.} {\bf 119} (2017), no.~25 251302,
  \href{http://xxx.lanl.gov/abs/1710.05877}{{\tt 1710.05877}}.

\bibitem{Ezquiaga:2017ekz}
J.~M. Ezquiaga and M.~Zumalacrregui, ``{Dark Energy After GW170817: Dead Ends
  and the Road Ahead},'' {\em Phys. Rev. Lett.} {\bf 119} (2017), no.~25
  251304, \href{http://xxx.lanl.gov/abs/1710.05901}{{\tt 1710.05901}}.

\bibitem{Baker:2017hug}
T.~Baker, E.~Bellini, P.~G. Ferreira, M.~Lagos, J.~Noller, and I.~Sawicki,
  ``{Strong constraints on cosmological gravity from GW170817 and GRB
  170817A},'' {\em Phys. Rev. Lett.} {\bf 119} (2017), no.~25 251301,
  \href{http://xxx.lanl.gov/abs/1710.06394}{{\tt 1710.06394}}.

\bibitem{Sakstein:2017xjx}
J.~Sakstein and B.~Jain, ``{Implications of the Neutron Star Merger GW170817
  for Cosmological Scalar-Tensor Theories},'' {\em Phys. Rev. Lett.} {\bf 119}
  (2017), no.~25 251303, \href{http://xxx.lanl.gov/abs/1710.05893}{{\tt
  1710.05893}}.

\bibitem{Crisostomi:2017lbg}
M.~Crisostomi and K.~Koyama, ``{Vainshtein mechanism after GW170817},'' {\em
  Phys. Rev.} {\bf D97} (2018), no.~2 021301,
  \href{http://xxx.lanl.gov/abs/1711.06661}{{\tt 1711.06661}}.

\bibitem{Langlois:2017dyl}
D.~Langlois, R.~Saito, D.~Yamauchi, and K.~Noui, ``{Scalar-tensor theories and
  modified gravity in the wake of GW170817},'' {\em Phys. Rev.} {\bf D97}
  (2018), no.~6 061501, \href{http://xxx.lanl.gov/abs/1711.07403}{{\tt
  1711.07403}}.

\bibitem{dima_vainshtein_2018}
A.~Dima and F.~Vernizzi, ``Vainshtein screening in scalar-tensor theories
  before and after {GW}170817: {Constraints} on theories beyond {Horndeski},''
  {\em Physical Review D} {\bf 97} (May, 2018).

\bibitem{Deffayet:2007kf}
C.~Deffayet and K.~Menou, ``{Probing Gravity with Spacetime Sirens},'' {\em
  Astrophys. J.} {\bf 668} (2007) L143--L146,
  \href{http://xxx.lanl.gov/abs/0709.0003}{{\tt 0709.0003}}.

\bibitem{Calabrese:2016bnu}
E.~Calabrese, N.~Battaglia, and D.~N. Spergel, ``{Testing Gravity with
  Gravitational Wave Source Counts},'' {\em Class. Quant. Grav.} {\bf 33}
  (2016), no.~16 165004, \href{http://xxx.lanl.gov/abs/1602.03883}{{\tt
  1602.03883}}.

\bibitem{Visinelli:2017bny}
L.~Visinelli, N.~Bolis, and S.~Vagnozzi, ``{Brane-world extra dimensions in
  light of GW170817},'' {\em Phys. Rev.} {\bf D97} (2018), no.~6 064039,
  \href{http://xxx.lanl.gov/abs/1711.06628}{{\tt 1711.06628}}.

\bibitem{Amendola:2017ovw}
L.~Amendola, I.~Sawicki, M.~Kunz, and I.~D. Saltas, ``{Direct detection of
  gravitational waves can measure the time variation of the Planck mass},''
  {\em JCAP} {\bf 1808} (2018), no.~08 030,
  \href{http://xxx.lanl.gov/abs/1712.08623}{{\tt 1712.08623}}.

\bibitem{Belgacem:2017ihm}
E.~Belgacem, Y.~Dirian, S.~Foffa, and M.~Maggiore, ``{Gravitational-wave
  luminosity distance in modified gravity theories},'' {\em Phys. Rev.} {\bf
  D97} (2018), no.~10 104066, \href{http://xxx.lanl.gov/abs/1712.08108}{{\tt
  1712.08108}}.

\bibitem{Creminelli:2018xsv}
P.~Creminelli, M.~Lewandowski, G.~Tambalo, and F.~Vernizzi, ``{Gravitational
  Wave Decay into Dark Energy},'' {\em JCAP} {\bf 1812} (2018), no.~12 025,
  \href{http://xxx.lanl.gov/abs/1809.03484}{{\tt 1809.03484}}.

\bibitem{Ezquiaga:2018btd}
J.~M. Ezquiaga and M.~Zumalacrregui, ``{Dark Energy in light of Multi-Messenger
  Gravitational-Wave astronomy},'' {\em Front. Astron. Space Sci.} {\bf 5}
  (2018) 44, \href{http://xxx.lanl.gov/abs/1807.09241}{{\tt 1807.09241}}.

\bibitem{Blanchet:2013haa}
L.~Blanchet, ``{Gravitational Radiation from Post-Newtonian Sources and
  Inspiralling Compact Binaries},'' {\em Living Rev. Rel.} {\bf 17} (2014) 2,
  \href{http://xxx.lanl.gov/abs/1310.1528}{{\tt 1310.1528}}.

\bibitem{goldberger_effective_2006}
W.~D. Goldberger and I.~Z. Rothstein, ``An {Effective} {Field} {Theory} of
  {Gravity} for {Extended} {Objects},'' {\em Physical Review D} {\bf 73} (May,
  2006). arXiv: hep-th/0409156.

\bibitem{Goldberger:2007hy}
W.~D. Goldberger, ``{Les Houches lectures on effective field theories and
  gravitational radiation},'' in {\em {Les Houches Summer School - Session 86:
  Particle Physics and Cosmology: The Fabric of Spacetime Les Houches, France,
  July 31-August 25, 2006}}, 2007.
\newblock \href{http://xxx.lanl.gov/abs/hep-ph/0701129}{{\tt hep-ph/0701129}}.

\bibitem{Cardoso:2008gn}
V.~Cardoso, O.~J.~C. Dias, and P.~Figueras, ``{Gravitational radiation in d>4
  from effective field theory},'' {\em Phys. Rev.} {\bf D78} (2008) 105010,
  \href{http://xxx.lanl.gov/abs/0807.2261}{{\tt 0807.2261}}.

\bibitem{Galley:2009px}
C.~R. Galley and M.~Tiglio, ``{Radiation reaction and gravitational waves in
  the effective field theory approach},'' {\em Phys. Rev.} {\bf D79} (2009)
  124027, \href{http://xxx.lanl.gov/abs/0903.1122}{{\tt 0903.1122}}.

\bibitem{Foffa:2013qca}
S.~Foffa and R.~Sturani, ``{Effective field theory methods to model compact
  binaries},'' {\em Class. Quant. Grav.} {\bf 31} (2014), no.~4 043001,
  \href{http://xxx.lanl.gov/abs/1309.3474}{{\tt 1309.3474}}.

\bibitem{porto_effective_2016}
R.~A. Porto, ``The {Effective} {Field} {Theorist}'s {Approach} to
  {Gravitational} {Dynamics},'' {\em Physics Reports} {\bf 633} (May, 2016)
  1--104. arXiv: 1601.04914.

\bibitem{Levi:2018nxp}
M.~Levi, ``{Effective Field Theories of Post-Newtonian Gravity: A comprehensive
  review},'' \href{http://xxx.lanl.gov/abs/1807.01699}{{\tt 1807.01699}}.

\bibitem{Endlich:2017tqa}
S.~Endlich, V.~Gorbenko, J.~Huang, and L.~Senatore, ``{An effective formalism
  for testing extensions to General Relativity with gravitational waves},''
  {\em JHEP} {\bf 09} (2017) 122,
  \href{http://xxx.lanl.gov/abs/1704.01590}{{\tt 1704.01590}}.

\bibitem{Nordtvedt:1968qs}
K.~Nordtvedt, ``{Equivalence Principle for Massive Bodies. 2. Theory},'' {\em
  Phys. Rev.} {\bf 169} (1968) 1017--1025.

\bibitem{damour_tensor-multi-scalar_1992}
T.~Damour and G.~Esposito-Farese, ``Tensor-multi-scalar theories of
  gravitation,'' {\em Classical and Quantum Gravity} {\bf 9} (Sept., 1992)
  2093--2176.

\bibitem{Damour:1995kt}
T.~Damour and G.~Esposito-Farese, ``{Testing gravity to second postNewtonian
  order: A Field theory approach},'' {\em Phys. Rev.} {\bf D53} (1996)
  5541--5578, \href{http://xxx.lanl.gov/abs/gr-qc/9506063}{{\tt
  gr-qc/9506063}}.

\bibitem{Huang:2018pbu}
J.~Huang, M.~C. Johnson, L.~Sagunski, M.~Sakellariadou, and J.~Zhang,
  ``{Prospects for axion searches with Advanced LIGO through binary mergers},''
  \href{http://xxx.lanl.gov/abs/1807.02133}{{\tt 1807.02133}}.

\bibitem{Julie:2017ucp}
F.-L. Juli, ``{Reducing the two-body problem in scalar-tensor theories to the
  motion of a test particle : a scalar-tensor effective-one-body approach},''
  {\em Phys. Rev.} {\bf D97} (2018), no.~2 024047,
  \href{http://xxx.lanl.gov/abs/1709.09742}{{\tt 1709.09742}}.

\bibitem{Julie:2017pkb}
F.-L. Juli and N.~Deruelle, ``{Two-body problem in Scalar-Tensor theories as a
  deformation of General Relativity : an Effective-One-Body approach},'' {\em
  Phys. Rev.} {\bf D95} (2017), no.~12 124054,
  \href{http://xxx.lanl.gov/abs/1703.05360}{{\tt 1703.05360}}.

\bibitem{Bernard:2018hta}
L.~Bernard, ``{Dynamics of compact binary systems in scalar-tensor theories:
  Equations of motion to the third post-Newtonian order},'' {\em Phys. Rev.}
  {\bf D98} (2018), no.~4 044004,
  \href{http://xxx.lanl.gov/abs/1802.10201}{{\tt 1802.10201}}.

\bibitem{Bernard:2018ivi}
L.~Bernard, ``{Dynamics of compact binary systems in scalar-tensor theories:
  II. Center-of-mass and conserved quantities to 3PN order},''
  \href{http://xxx.lanl.gov/abs/1812.04169}{{\tt 1812.04169}}.

\bibitem{Lang:2013fna}
R.~N. Lang, ``{Compact binary systems in scalar-tensor gravity. II. Tensor
  gravitational waves to second post-Newtonian order},'' {\em Phys. Rev.} {\bf
  D89} (2014), no.~8 084014, \href{http://xxx.lanl.gov/abs/1310.3320}{{\tt
  1310.3320}}.

\bibitem{Lang:2014osa}
R.~N. Lang, ``{Compact binary systems in scalar-tensor gravity. III. Scalar
  waves and energy flux},'' {\em Phys. Rev.} {\bf D91} (2015), no.~8 084027,
  \href{http://xxx.lanl.gov/abs/1411.3073}{{\tt 1411.3073}}.

\bibitem{Eardley1975ApJ}
D.~M. {Eardley}, ``{Observable effects of a scalar gravitational field in a
  binary pulsar},'' {\em Astrophysical Journal} {\bf 196} (Mar., 1975)
  L59--L62.

\bibitem{Nicolis:2008in}
A.~Nicolis, R.~Rattazzi, and E.~Trincherini, ``{The Galileon as a local
  modification of gravity},'' {\em Phys. Rev.} {\bf D79} (2009) 064036,
  \href{http://xxx.lanl.gov/abs/0811.2197}{{\tt 0811.2197}}.

\bibitem{Horndeski:1974wa}
G.~W. Horndeski, ``{Second-order scalar-tensor field equations in a
  four-dimensional space},'' {\em Int. J. Theor. Phys.} {\bf 10} (1974)
  363--384.

\bibitem{Deffayet:2009mn}
C.~Deffayet, S.~Deser, and G.~Esposito-Farese, ``{Generalized Galileons: All
  scalar models whose curved background extensions maintain second-order field
  equations and stress-tensors},'' {\em Phys. Rev.} {\bf D80} (2009) 064015,
  \href{http://xxx.lanl.gov/abs/0906.1967}{{\tt 0906.1967}}.

\bibitem{Zumalacarregui:2013pma}
M.~Zumalacrregui and J.~Garca-Bellido, ``{Transforming gravity: from derivative
  couplings to matter to second-order scalar-tensor theories beyond the
  Horndeski Lagrangian},'' {\em Phys. Rev.} {\bf D89} (2014) 064046,
  \href{http://xxx.lanl.gov/abs/1308.4685}{{\tt 1308.4685}}.

\bibitem{Gleyzes:2014dya}
J.~Gleyzes, D.~Langlois, F.~Piazza, and F.~Vernizzi, ``{Healthy theories beyond
  Horndeski},'' {\em Phys. Rev. Lett.} {\bf 114} (2015), no.~21 211101,
  \href{http://xxx.lanl.gov/abs/1404.6495}{{\tt 1404.6495}}.

\bibitem{Gleyzes:2014qga}
J.~Gleyzes, D.~Langlois, F.~Piazza, and F.~Vernizzi, ``{Exploring gravitational
  theories beyond Horndeski},'' {\em JCAP} {\bf 1502} (2015) 018,
  \href{http://xxx.lanl.gov/abs/1408.1952}{{\tt 1408.1952}}.

\bibitem{Langlois:2015cwa}
D.~Langlois and K.~Noui, ``{Degenerate higher derivative theories beyond
  Horndeski: evading the Ostrogradski instability},'' {\em JCAP} {\bf 1602}
  (2016), no.~02 034, \href{http://xxx.lanl.gov/abs/1510.06930}{{\tt
  1510.06930}}.

\bibitem{BenAchour:2016fzp}
J.~Ben~Achour, M.~Crisostomi, K.~Koyama, D.~Langlois, K.~Noui, and G.~Tasinato,
  ``{Degenerate higher order scalar-tensor theories beyond Horndeski up to
  cubic order},'' {\em JHEP} {\bf 12} (2016) 100,
  \href{http://xxx.lanl.gov/abs/1608.08135}{{\tt 1608.08135}}.

\bibitem{Vainshtein:1972sx}
A.~I. Vainshtein, ``{To the problem of nonvanishing gravitation mass},'' {\em
  Phys. Lett.} {\bf 39B} (1972) 393--394.

\bibitem{Babichev:2013usa}
E.~Babichev and C.~Deffayet, ``{An introduction to the Vainshtein mechanism},''
  {\em Class. Quant. Grav.} {\bf 30} (2013) 184001,
  \href{http://xxx.lanl.gov/abs/1304.7240}{{\tt 1304.7240}}.

\bibitem{Creminelli:2006xe}
P.~Creminelli, M.~A. Luty, A.~Nicolis, and L.~Senatore, ``{Starting the
  Universe: Stable Violation of the Null Energy Condition and Non-standard
  Cosmologies},'' {\em JHEP} {\bf 12} (2006) 080,
  \href{http://xxx.lanl.gov/abs/hep-th/0606090}{{\tt hep-th/0606090}}.

\bibitem{Cheung:2007st}
C.~Cheung, P.~Creminelli, A.~L. Fitzpatrick, J.~Kaplan, and L.~Senatore, ``{The
  Effective Field Theory of Inflation},'' {\em JHEP} {\bf 03} (2008) 014,
  \href{http://xxx.lanl.gov/abs/0709.0293}{{\tt 0709.0293}}.

\bibitem{Creminelli:2008wc}
P.~Creminelli, G.~D'Amico, J.~Norena, and F.~Vernizzi, ``{The Effective Theory
  of Quintessence: the w<-1 Side Unveiled},'' {\em JCAP} {\bf 0902} (2009) 018,
  \href{http://xxx.lanl.gov/abs/0811.0827}{{\tt 0811.0827}}.

\bibitem{Gubitosi:2012hu}
G.~Gubitosi, F.~Piazza, and F.~Vernizzi, ``{The Effective Field Theory of Dark
  Energy},'' {\em JCAP} {\bf 1302} (2013) 032,
  \href{http://xxx.lanl.gov/abs/1210.0201}{{\tt 1210.0201}}.
  [JCAP1302,032(2013)].

\bibitem{Bloomfield:2012ff}
J.~K. Bloomfield, a.~. Flanagan, M.~Park, and S.~Watson, ``{Dark energy or
  modified gravity? An effective field theory approach},'' {\em JCAP} {\bf
  1308} (2013) 010, \href{http://xxx.lanl.gov/abs/1211.7054}{{\tt 1211.7054}}.

\bibitem{Piazza:2013coa}
F.~Piazza and F.~Vernizzi, ``{Effective Field Theory of Cosmological
  Perturbations},'' {\em Class. Quant. Grav.} {\bf 30} (2013) 214007,
  \href{http://xxx.lanl.gov/abs/1307.4350}{{\tt 1307.4350}}.

\bibitem{Babichev:2009ee}
E.~Babichev, C.~Deffayet, and R.~Ziour, ``{k-Mouflage gravity},'' {\em Int. J.
  Mod. Phys.} {\bf D18} (2009) 2147--2154,
  \href{http://xxx.lanl.gov/abs/0905.2943}{{\tt 0905.2943}}.

\bibitem{Chu:2012kz}
Y.-Z. Chu and M.~Trodden, ``{Retarded Greens function of a Vainshtein system
  and Galileon waves},'' {\em Phys. Rev.} {\bf D87} (2013), no.~2 024011,
  \href{http://xxx.lanl.gov/abs/1210.6651}{{\tt 1210.6651}}.

\bibitem{deRham:2012fw}
C.~de~Rham, A.~J. Tolley, and D.~H. Wesley, ``{Vainshtein Mechanism in Binary
  Pulsars},'' {\em Phys. Rev.} {\bf D87} (2013), no.~4 044025,
  \href{http://xxx.lanl.gov/abs/1208.0580}{{\tt 1208.0580}}.

\bibitem{deRham:2012fg}
C.~de~Rham, A.~Matas, and A.~J. Tolley, ``{Galileon Radiation from Binary
  Systems},'' {\em Phys. Rev.} {\bf D87} (2013), no.~6 064024,
  \href{http://xxx.lanl.gov/abs/1212.5212}{{\tt 1212.5212}}.

\bibitem{Dar:2018dra}
F.~Dar, C.~De~Rham, J.~T. Deskins, J.~T. Giblin, and A.~J. Tolley, ``{Scalar
  Gravitational Radiation from Binaries: Vainshtein Mechanism in Time-dependent
  Systems},'' \href{http://xxx.lanl.gov/abs/1808.02165}{{\tt 1808.02165}}.

\bibitem{Weinberg:1964ew}
S.~Weinberg, ``{Photons and Gravitons in s Matrix Theory: Derivation of Charge
  Conservation and Equality of Gravitational and Inertial Mass},'' {\em Phys.
  Rev.} {\bf 135} (1964) B1049--B1056.

\bibitem{Hui:2010dn}
L.~Hui and A.~Nicolis, ``{An Equivalence principle for scalar forces},'' {\em
  Phys. Rev. Lett.} {\bf 105} (2010) 231101,
  \href{http://xxx.lanl.gov/abs/1009.2520}{{\tt 1009.2520}}.

\bibitem{Nitti:2012ev}
F.~Nitti and F.~Piazza, ``{Scalar-tensor theories, trace anomalies and the
  QCD-frame},'' {\em Phys. Rev.} {\bf D86} (2012) 122002,
  \href{http://xxx.lanl.gov/abs/1202.2105}{{\tt 1202.2105}}.

\bibitem{Hui:2009kc}
L.~Hui, A.~Nicolis, and C.~Stubbs, ``{Equivalence Principle Implications of
  Modified Gravity Models},'' {\em Phys. Rev.} {\bf D80} (2009) 104002,
  \href{http://xxx.lanl.gov/abs/0905.2966}{{\tt 0905.2966}}.

\bibitem{Will:1993ns}
C.~M. Will, {\em {Theory and experiment in gravitational physics}}.
\newblock 1993.

\bibitem{einstein_gravitational_1938}
A.~Einstein, L.~Infeld, and B.~Hoffmann, ``The {Gravitational} {Equations} and
  the {Problem} of {Motion},'' {\em The Annals of Mathematics} {\bf 39} (Jan.,
  1938) 65.

\bibitem{Kol:2007bc}
B.~Kol and M.~Smolkin, ``{Non-Relativistic Gravitation: From Newton to Einstein
  and Back},'' {\em Class. Quant. Grav.} {\bf 25} (2008) 145011,
  \href{http://xxx.lanl.gov/abs/0712.4116}{{\tt 0712.4116}}.

\bibitem{goldberger_gravitational_2010}
W.~D. Goldberger and A.~Ross, ``Gravitational radiative corrections from
  effective field theory,'' {\em Physical Review D} {\bf 81} (June, 2010).
  arXiv: 0912.4254.

\bibitem{ross_multipole_2012}
A.~Ross, ``Multipole expansion at the level of the action,'' {\em Physical
  Review D} {\bf 85} (June, 2012). arXiv: 1202.4750.

\bibitem{galley_radiation_2009}
C.~R. Galley and M.~Tiglio, ``Radiation reaction and gravitational waves in the
  effective field theory approach,'' {\em Physical Review D} {\bf 79} (June,
  2009).

\bibitem{Cardoso:2002pa}
V.~Cardoso, O.~J.~C. Dias, and J.~P.~S. Lemos, ``{Gravitational radiation in
  D-dimensional space-times},'' {\em Phys. Rev.} {\bf D67} (2003) 064026,
  \href{http://xxx.lanl.gov/abs/hep-th/0212168}{{\tt hep-th/0212168}}.

\bibitem{Maggiore:1900zz}
M.~Maggiore, {\em {Gravitational Waves. Vol. 1: Theory and Experiments}}.
\newblock Oxford Master Series in Physics. Oxford University Press, 2007.

\bibitem{Will:1994fb}
C.~M. Will, ``{Testing scalar - tensor gravity with gravitational wave
  observations of inspiraling compact binaries},'' {\em Phys. Rev.} {\bf D50}
  (1994) 6058--6067, \href{http://xxx.lanl.gov/abs/gr-qc/9406022}{{\tt
  gr-qc/9406022}}.

\bibitem{Sennett:2016klh}
N.~Sennett, S.~Marsat, and A.~Buonanno, ``{Gravitational waveforms in
  scalar-tensor gravity at 2PN relative order},'' {\em Phys. Rev.} {\bf D94}
  (2016), no.~8 084003, \href{http://xxx.lanl.gov/abs/1607.01420}{{\tt
  1607.01420}}.

\bibitem{Gleyzes:2015pma}
J.~Gleyzes, D.~Langlois, M.~Mancarella, and F.~Vernizzi, ``{Effective Theory of
  Interacting Dark Energy},'' {\em JCAP} {\bf 1508} (2015), no.~08 054,
  \href{http://xxx.lanl.gov/abs/1504.05481}{{\tt 1504.05481}}.

\bibitem{Brax:2018bow}
P.~Brax and A.-C. Davis, ``{Gravitational effects of disformal couplings},''
  {\em Phys. Rev.} {\bf D98} (2018), no.~6 063531,
  \href{http://xxx.lanl.gov/abs/1809.09844}{{\tt 1809.09844}}.

\bibitem{Inprep}
A.~Kuntz, P.~Brax, and A.-C. Davis in prep.

\bibitem{Will:2014kxa}
C.~M. Will, ``{The Confrontation between General Relativity and Experiment},''
  {\em Living Rev. Rel.} {\bf 17} (2014) 4,
  \href{http://xxx.lanl.gov/abs/1403.7377}{{\tt 1403.7377}}.

\bibitem{Bertotti:2003rm}
B.~Bertotti, L.~Iess, and P.~Tortora, ``{A test of general relativity using
  radio links with the Cassini spacecraft},'' {\em Nature} {\bf 425} (2003)
  374--376.

\end{thebibliography}\endgroup

\end{document}